\documentclass[twocolumn,amsmath,amssymb,superscriptaddress,pra,a4paper]{revtex4-1}
\usepackage{graphicx}
\usepackage[paperwidth=212mm,paperheight=297mm,centering,hmargin=1.65cm,vmargin=2.6cm]{geometry}
\usepackage{braket}
\usepackage{bbold}
\graphicspath{{graphics/}}
\newcommand{\e}{\mathrm{e}}
\newcommand{\ketbra}[2]{\left|#1\right\rangle\hskip-1mm\left\langle #2\right|}

\begin{document}

\title{Stabilisation of an optical transition energy via nuclear Zeno dynamics in \\ quantum dot-cavity systems}
\author{Thomas Nutz}
\email{nutzat@gmail.com}
\affiliation{Controlled Quantum Dynamics Theory Group, Imperial College London, London SW7 2AZ, United Kingdom}
\affiliation{Quantum Engineering Technology Labs, H. H. Wills Physics Laboratory and Department of Electrical and Electronic Engineering, 
University of Bristol, BS8 1FD, UK}
\author{Petros Androvitsaneas}
\author{Andrew Young}
\author{Ruth Oulton}
\author{Dara P. S. McCutcheon}
\affiliation{Quantum Engineering Technology Labs, H. H. Wills Physics Laboratory and Department of Electrical and Electronic Engineering, 
University of Bristol, BS8 1FD, UK}

\date{\today}

\begin{abstract}
We investigate the effect of nuclear spins on the phase shift and polarisation rotation of photons scattered off a 
quantum dot--cavity system. We show that as the phase shift depends strongly on the resonance energy of an 
electronic transition in the quantum dot, it can provide a sensitive probe of the quantum state of nuclear spins 
that broaden this transition energy. By including the electron--nuclear spin coupling at a Hamiltonian level within an extended input--output formalism, we show how a photon scattering event acts as a nuclear spin measurement, which when rapidly applied leads to an inhibition of the nuclear dynamics via the quantum Zeno effect, and a corresponding stabilisation of the optical resonance. We show how such an effect manifests in the 
intensity autocorrelation $g^{(2)}(\tau)$ of scattered photons, whose long-time bunching behaviour changes from quadratic decay for low photon scattering rates (weak laser intensities), to ever slower exponential decay for increasing laser intensities as optical measurements impede the nuclear spin evolution.
\end{abstract}

\maketitle

\section{Introduction}
The generation of useful entanglement between photons is the central challenge in optical quantum computing schemes.
Self-assembled quantum dots (QDs) have the potential to meet this challenge, either by emitting strings of entangled 
photons~\cite{lindner_proposal_2009, economou_optically_2010, schwartz_deterministic_2016-2}, or by mediating an effective 
interaction between photons via a giant phase shift~\cite{hu_giant_2008, hu_deterministic_2008, pineiro-orioli_noise_2013, lodahl_interfacing_2015, lodahl_quantum-dot_2018}. 
Current experimental efforts to utilise such schemes, however, are often hindered by noise arising due to the coupling of an electron spin to the nuclear spins in the host material~\cite{mccutcheon_error_2014, kuhlmann_transform-limited_2015, stockill_quantum_2016, wuest_role_2016, ethier-majcher_improving_2017}. Nevertheless, the dephasing caused by these nuclear spins is qualitatively different from 
that caused by coupling to photon or phonon baths, 
as the nuclear spins evolve slowly and unitarily on the timescale 
set by the electron spin dynamics, which 
gives rise to a variety of non-Markovian effects~\cite{greilich_nuclei-induced_2007, barnes_electron-nuclear_2011, madsen_observation_2011, urbaszek_nuclear_2013, economou_theory_2014, munsch_manipulation_2014, prechtel_decoupling_2016, ethier-majcher_improving_2017, nutz_solvable_2019}. While this unique nature of the nuclear spin environment might make it possible to experimentally suppress nuclear spin noise and possibly even control them in a useful way, it also presents a formidable theoretical challenge 
to find reliable and insightful models of nuclear spin behaviour.

\begin{figure*}
\includegraphics[width=0.9\textwidth]{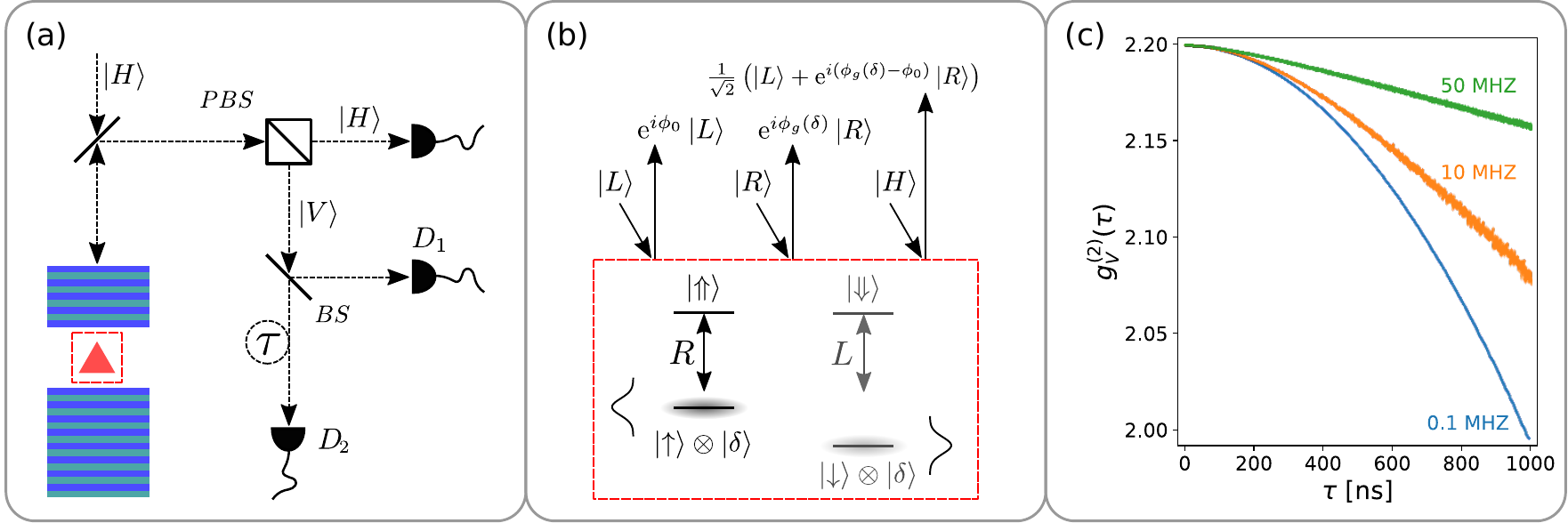}
\caption{
(a) Experimental setup used to measure photon phase shifts due to a quantum dot inside a micropillar cavity. 
The states $\ket{H}$ and $\ket{V}$ denote horizontally and vertically polarised light, while (P)BS labels a (polarising) 50/50 beamsplitter. 
A variable delay $\tau$ between detectors $D_1$ and $D_2$ can be used to measure the 
intensity autocorrelation function of light scattered into cross polarised (vertical) channel, as shown in (c). 
(b) Right $\ket{R}$ and left $\ket{L}$ circularly polarised photons couple to the spin ground states 
$\ket{\uparrow}$ and $\ket{\downarrow}$ respectively. In a large magnetic field, the $\ket{\downarrow}$ is 
far detuned, and $\ket{L}$ reflects off an effectively empty cavity, while $\ket{R}$ experiences 
a phase shift $\phi_g(\delta)$ that depends on the nuclear spin Overhauser field $\delta$. 
(c) Cross-polarised intensity 
correlation function for different average photon count rates obtained from a Monte-Carlo simulation 
including 8 nuclear spins. 
While the quadratic behaviour characteristic of unitary evolution is observed for low count rates, 
the evolution of the nuclear spins is impeded by more frequent photon scattering, which constitutes a quantum Zeno effect. Note that the antibunching at sub-nanosecond timescales is neglected and that the weak-driving assumption is satisfied.
Parameters $A_k$ and $\omega_k$ were randomly drawn from a Gaussian distribution with 
$\braket{A_k} = \braket{\omega_k}$ = 0.5 $\mu$eV, $\sigma (A_k) = 0.25\mu$eV, and $\sigma(\omega_k) = 10$neV.
}
 \label{fig:Zeno}
\end{figure*}

We consider the effect of nuclear spins in giant phase shift experiments such as those described in Ref.~\cite{androvitsaneas_charged_2016} 
(see Fig. \ref{fig:Zeno}a), in which narrowband laser photons of linear polarisation described by $\ket{H}\propto\ket{L}+\ket{R}$ scatter off a cavity containing a 
charged QD in a large ($\gtrsim 100$ mT) magnetic field in the Faraday configuration. Since an electronic transition couples only to one of the two circular polarisations $\ket{L}$ and $\ket{R}$, the photon polarisation state upon scattering is given by $\e^{i \theta_L}\ket{L}+\e^{i \theta_R}\ket{R}$, with the phase shift difference $\theta = \theta_L-\theta_R$ taking values of up to $180^{\circ}$ \cite{hu_giant_2008, hu_deterministic_2008,auffeves-garnier_giant_2007}. Hence, a linearly polarised photon $\ket{H}$ can be reflected with the orthogonal linear polarisation $\ket{V}\propto \ket{L}-\ket{R}$, as shown in Fig. \ref{fig:Zeno}b. The phase shift is highly sensitive to the resonance energy of the electronic transition, which in turn depends on the nuclear spin environment via the Overhauser shift \cite{abragam_principles_1961}. 
As the nuclear spin system evolves, the phase shift $\theta$ drifts over time, 
such that high values are observed only during short intervals ($\theta > 120^{\circ}$ in $100 \mu\mathrm{s}$ timebins~\cite{androvitsaneas_efficient_2019}) but the time-averaged phase 
shift is low ($\braket{\theta}  \approx 6^{\circ}$ in \cite{androvitsaneas_efficient_2019}). Photon detection events in the cross polarised (orthogonal to input laser) channel are therefore bunched on a $\mu s$ timescale, such that an intensity autocorrelation function has $g^{(2)}(\tau) < 1$ for 
$\tau < \mathrm{ns}$ due to the single photon nature of the scattered field, but $g^{(2)}(\tau) > 1$ for $\tau\sim \mu\mathrm{s}$ 
as the nuclear spin coupling effectively leads to blinking.

In this work we develop a quantum optical treatment that relates the intensity correlation function in the cross polarised channel $g^{(2)}(\tau)$ to a two-time correlation function of the nuclear spin system. We show that $g^{(2)}(\tau)$ decreases 
quadratically for low laser intensities, as depicted by the blue 
curve in Fig. \ref{fig:Zeno}c. Observation of this quadratic short-time behaviour would demonstrate the coherent nature of nuclear spin noise in QDs, 
which could help distinguish it from other possible sources of resonance fluctuations in these systems. 
However, the dependence of the photon phase shift on the nuclear spin state is only one aspect of 
a two-way interaction, as a photon scattering event has the effect of a quantum measurement on the 
nuclear spin state. Incorporating this into our formalism, we find that 
frequent photon scattering events, corresponding to higher driving intensities, 
impede the nuclear spin evolution and associated drifting of the resonance energy, which leads to a broadened intensity autocorrelation function that decays linearly with $\tau$. This can be 
understood as a quantum Zeno effect~\cite{misra_zenos_1977, itano_quantum_1990, block_quantum_1991, 
facchi_quantum_2000, pascazio_all_2014, zhang_zeno_2015, christensen_driving-induced_2018}, 
which is here readily observable in an optical intensity correlation function.
Experimental observation of this characteristic change in the intensity 
autocorrelation function would demonstrate this novel quantum 
Zeno effect, and open up a measurement-based route to control nuclear spins in QDs.

\section{Input--output formalism with electron--nuclear spin coupling}

Our aim is to calculate the cross-polarised intensity autocorrelation $g^{(2)}(\tau)$ for photons scattered 
off the QD-cavity system which incorporates the 
nuclear spin environment, and which we achieve using an extended input-output 
formalism~\cite{gardiner_input_1985, walls_quantum_2008, auffeves-garnier_giant_2007}. 
We consider a continuum of optical modes described by annihilation operators 
$\smash{{b}(\omega)}$
propagating 
towards and away from an optical cavity with frequency $\omega_c$ and associated cavity 
mode operator 
${a}$. The cavity mode, in turn, couples to a two-level system (TLS) with ground and 
excited states $\ket{\uparrow}$ and $\ket{\Uparrow}$, respectively, which itself 
is coupled to a bath of nuclear spins. 
The total Hamiltonian describing all degrees of freedom 
is written $H = H_0 + H_{I}$, with (setting $\hbar=1$) 
\begin{align}
H_0 &= \frac{1}{2}\omega_0 \sigma^z + \omega_c {a}^{\dagger} {a} 
+ \int_0^{\infty}\!\!\!d\omega\, \omega {b}^{\dagger}(\omega) {b}(\omega) + H_Z,  \\
H_{I} &= g \left( \sigma^- {a}^{\dagger} + \sigma^+ {a} \right) \nonumber \\
&+\int_0^{\infty} \!\!\!d\omega \sqrt{\kappa(\omega)} \left( {b}(\omega) {a}^{\dagger} + {b}^{\dagger}(\omega) {a} \right) + H_{O},
 \label{eq:free and interacting Hamiltonian compononents}
\end{align}
where $\sigma^z=\ketbra{\Uparrow}{\Uparrow}-\ketbra{\uparrow}{\uparrow}$, $\sigma^-=\ketbra{\uparrow}{\Uparrow}$, 
$\sigma^+=\ketbra{\Uparrow}{\uparrow}$, $\omega_0$ is the transition energy of the TLS, 
and $g$ is the TLS--cavity coupling strength. The nuclear Zeeman term is $H_Z = \sum_j \omega_j I_j^z$, with Pauli $z$ operator $I^z_j$ 
acting on nuclear spin $j$ and nuclear Zeeman splitting $\omega_j$ due to 
an external magnetic field along $\hat{z}$. 
The electron--nuclear coupling term is $H_{O}=\frac{1}{2}\sigma^z \hat{\Delta}$, with Overhauser shift operator 
\begin{equation}
\hat{\Delta} = \sum_j A_j I_j^z + \frac{1}{2 \omega} \sum_{m\neq n} A_m A_n I^+_m I^-_n, 
 \label{eq:Delta}
\end{equation}
which results from a Schrieffer-Wolff transformation on the contact hyperfine 
Hamiltonian given in Eq. \ref{eq:contact hyperfine Hamiltonian} ~\cite{klauser_nuclear_2008}. Note that while the contact hyperfine interaction 
involves two electron spin states, $\ket{\uparrow}$ and $\ket{\downarrow}$, we focus here on 
one of these ground states only, arbitrarily labelled $\ket{\uparrow}$. Neglecting the other spin state  
$\ket{\downarrow}$ is justified in a large ($\gtrsim 100 $ mT) magnetic field, where energy 
conservation prevents flip-flops between these electron spin states which are separated by the 
electron Zeeman energy $\omega$. 

We approximate the cavity--port mode coupling 
strength as a constant $\sqrt{\kappa(\omega)} \approx \sqrt{\kappa(\omega_c)} \equiv \sqrt{\kappa}$ over the relevant optical frequencies, 
and in doing so we find the Heisenberg equations of motion 
\begin{subequations}
\begin{align}
\dot{\sigma}^-(t) &\!= i g\ {a}(t) \sigma_z(t) -i\Big(\omega_0 + \hat{\Delta}(t)\Big)\sigma^-(t), \label{Ha}\\
i\ \dot{a}(t) &\!= (\omega_c - i \pi \kappa) a(t) + g \sigma^-(t) + \sqrt{\kappa}\ b_{\mathrm{in}}(t), \label{Hb}\\
\!\!\!2 \pi i \sqrt{\kappa}\ {a}(t) &\!= {b}_{\mathrm{in}}(t) - {b}_{\mathrm{out}}(t), \label{Hc}
\end{align}
\end{subequations}
where we have defined the incoming and outgoing field operators as 
${b}_{\mathrm{in}}(t) = \int d \omega \  b_0(\omega)\exp (-i \omega t)$ with 
$b_0(\omega) \equiv {b}(\omega, t_0)$, and ${b}_{\mathrm{out}}(t) = \int d \omega \  b_1(\omega)\exp (-i \omega t)$ 
with $b_1(\omega) \equiv {b}(\omega, t_1)$~\cite{walls_quantum_2008}, and 
extended frequency integrals such that 
$\int_0^{\infty}d \omega \rightarrow \int_{-\infty}^{\infty}d \omega \equiv \int d \omega$. 
Taking the Fourier transform of Eq.~({\ref{Hb}) we find 
\begin{equation}
\left( \omega - \omega_c + i \pi \kappa \right) a(\omega) =  g \sigma^- (\omega) + \sqrt{\kappa} b_0 (\omega), 
 \label{eq: a Fourier equality}
\end{equation}
where $\sigma^-(t)=\int d\omega \ \sigma^-(\omega)\e^{-i\omega t}$ and similarly for ${a}(t)$. 

The standard procedure in input--output theory is to use 
the Fourier transform of Eq.~({\ref{Ha}}) to 
replace $\sigma^-(\omega)$ in Eq.~({\ref{eq: a Fourier equality}}), which is then 
used in the Fourier transform of Eq.~({\ref{Hc}}), 
$b_{\mathrm{in}}(\omega) - b_{\mathrm{out}}(\omega) = 2 \pi i \sqrt{\kappa} a(\omega)$ 
to find a relationship between frequency components of the 
incoming and outgoing fields $b_{\mathrm{in}}(\omega)$ and $b_{\mathrm{out}}(\omega)$. 
We use a similar procedure here, but note that the 
occurrence of the time-dependent Overhauser shift operator $\hat{\Delta}(t)$ in Eq.~({\ref{Ha}}) 
means there is no simple relationship between the Fourier components $\sigma(\omega)$ and $a(\omega)$. 
Instead we arrive at the integral equation 
\begin{equation}
\int d\omega \left( [ \omega - \hat{\omega}_0(t) ] \sigma^-(\omega) - g a(\omega) \right) \e^{-i \omega t} = 0,
 \label{eq:sigma_minus Fourier Heisenberg}
\end{equation}
where we have defined $\hat{\omega}_0(t)\equiv \omega_0 + \hat{\Delta}(t)$ and 
made the approximation $\sigma_z \approx -1$, valid for weak driving. Combining Eqs. \eqref{eq: a Fourier equality} and \eqref{eq:sigma_minus Fourier Heisenberg} then gives 

\begin{equation}
\int \!\!d\omega \e^{- i \omega t} \hat{f}_+(\omega, t) a(\omega) = \!\!\int \!\!d\omega \e^{- i \omega t} \sqrt{\kappa}[ \omega - \hat{\omega}_0(t)]b_0(\omega),
\end{equation}
where $\hat{f}_{\pm}= \left( \omega - \omega_c \pm i \pi \kappa \right) \left( \omega - \hat{\omega}_0 (t) \right) - g^2$.
Using this in the Fourier transformation of Eq.~({\ref{Hc}}) leads to
\begin{equation}
\int d \omega \left( b_{\mathrm{out}}(\omega) \hat{f}_+(\omega,t) -  b_{\mathrm{in}}(\omega) \hat{f}_-(\omega,t)\right) \e^{-i\omega t} = 0.
 \label{eq:integral equation bout bin}
\end{equation}

When the Overhauser term is neglected, $\hat{\Delta}(t)=0$, 
Eq.~({\ref{eq:integral equation bout bin}}) 
simplifies to $\smash{{b}_{\mathrm{out}}(\omega) = r(\omega) b_{\mathrm{in}}(\omega)}$ with the scalar $r(\omega) = f_-(\omega) / f_+(\omega)$, 
which is the well-known cavity-QED reflection coefficient~\cite{auffeves-garnier_giant_2007}. 
An analogous result relating incoming and outgoing fields in the presence of nuclear spin coupling can be obtained by assuming 
a slowly varying Overhauser shift. To see this, we consider attempting to isolate the integrand 
in Eq.~({\ref{eq:integral equation bout bin}}) by performing the finite-domain definite integral 
\begin{equation}\begin{split}
&\int_{t_c - t_{\Delta}}^{t_c + t_{\Delta}}dt \,\e^{i \omega^{\prime}t} \int d \omega \  b_{\mathrm{out}}(\omega) \hat{f}_+(\omega,t) \e^{-i\omega t}\\
= &\int_{t_c - t_{\Delta}}^{t_c + t_{\Delta}}dt\, \e^{i \omega^{\prime}t} \int d \omega \  b_{\mathrm{in}}(\omega) \hat{f}_-(\omega,t) \e^{-i\omega t}.
\end{split} \label{eq:multiply exp and integrate}
\end{equation}
Choosing the integration range such that $t_{\Delta} \ll t_{\mathrm{fluc}}$, where 
$t_{\mathrm{fluc}}$ is the characteristic timescale of the Overhauser 
field fluctuations, we can approximate $\hat{\Delta}(t) \approx \hat{\Delta}(t_c)$ in the integrands and arrive at
\begin{equation}\begin{split}
&\int d \omega \  b_{\mathrm{out}}(\omega,t_c) \hat{f}_+(\omega,t_c) t_{\Delta} \mathrm{sinc} [t_{\Delta}(\omega - \omega^{\prime})] \\
= &\int d \omega \  b_{\mathrm{in}}(\omega,t_c) \hat{f}_-(\omega,t_c) t_{\Delta} \mathrm{sinc} [t_{\Delta}(\omega - \omega^{\prime})],
\end{split} \label{eq:integral equation bout bin 2}
\end{equation}
where $b_{\mathrm{out}}(\omega, t_c) \equiv b_{\mathrm{out}}(\omega) \e^{- i \omega t_c}$ and similarly for 
$b_{\mathrm{in}}(\omega,t_c)$. 
If the Overhauser shift fluctuates slowly then we can choose the spectral width of the 
sinc function in Eq.~({\ref{eq:integral equation bout bin 2}}), $t_{\Delta}^{-1}$, to be 
much narrower than the width of the function $\smash{\hat{f}_{\pm}(\omega,t_c)}$. We then find 
\begin{equation}
\tilde{b}_{\mathrm{out}}(\omega) =\hat{r}(\omega,t)\otimes \tilde{b}_{\mathrm{in}}(\omega),
 \label{eq:reflectivity operator photons}
\end{equation}
with $\hat{r}(\omega,t)=f_+(\omega,t)^{-1}f_-(\omega,t)$ and given by 
\begin{equation}\begin{split}
\hat{r}(\omega,t) = \mathbb{1} + \frac{2 i \pi \kappa (\omega_0 + \hat{\Delta}(t)-\omega)}{(\omega_c - \omega- i \pi \kappa )(\omega_0 + \hat{\Delta}(t)-\omega) - g^2},
\end{split}
 \label{eq:time dependent reflectivity}
\end{equation}
while the incoming and outgoing field operators are now defined as 
\begin{equation}\begin{split}
& \tilde{b}_{\mathrm{out}}(\omega) \equiv \int d\omega^{\prime}\ b_{\mathrm{out}}(\omega^{\prime},t)t_{\Delta}\mathrm{sinc}[(\omega -\omega^{\prime}) t_{\Delta}], 
\end{split}
 \label{eq:tilde b in/out}
\end{equation}
and similarly for $\tilde{b}_{\mathrm{in}}(\omega)$. 
Noticing the convolution form of this expression, we see that this operator 
can be thought of as a broadened version of its exact frequency 
counterpart $b_{\mathrm{out}}(\omega)$ owing to the finite 
integration time $t_\Delta$.

The relationship between incoming and outgoing fields given in 
Eq.~(\ref{eq:reflectivity operator photons}) is our first result, and generalises 
input--output theory to systems with slowly varying resonance energies. 
It is valid if the integration time $t_{\Delta}$ in Eq.~(\ref{eq:multiply exp and integrate}) 
satisfies $t_{\mathrm{fluc}} \gg t_{\Delta} \gg 1/w_f$, where 
$t_{\mathrm{fluc}}$ is the fluctuation time of the Overhauser shift and $w_f$ is the 
spectral width of the phase shift feature, which is obtained by considering the 
frequency dependence of $\hat{f}_{\pm}(\omega,t) = \hat{f}(\omega,t) \exp (\pm i \hat{\theta} (\omega,t))$. 
We find that the phase factor varies most rapidly at $\omega = \omega_c = \hat{\omega}_0$, at which point 
\begin{equation}
\frac{d }{d \omega} \hat{\theta}(\omega,t) = 2 \pi \frac{\kappa}{g^2},
 \label{eq:phase derivative}
\end{equation}
and $\frac{d}{d\omega} \hat{f}(\omega,t)=0$. As such, the fractional 
variation $\frac{d}{d\omega} \hat{f}(\omega,t)/\hat{f}(\omega,t)$ does not exceed a 
bound on the order of $\kappa / g^2$ when considering laser--QD detunings no greater 
than $ \omega - \hat{\omega}_0(t) \approx g^2/\kappa$, and laser--cavity detunings limited 
to $\omega - \omega_c \approx \kappa$. The functions $\smash{\hat{f}_{\pm}(\omega,t)}$ therefore 
vary on a frequency scale given by the linewidth $g^2 / \kappa$ of the TLS transition. 
This linewidth is typically on the order of few GHz for QD experiments, while the 
Overhauser shift fluctuation time can be estimated to be hundreds of milliseconds based 
on~\cite{androvitsaneas_efficient_2019}, such that $t_{\mathrm{fluc}} \gg t_{\Delta} \gg 1/w_f$ 
can be satisfied and Eq.~(\ref{eq:reflectivity operator photons}) is applicable to QD experiments. We interpret 
$t_{\Delta}$ as a parameter that adjusts the tradeoff between frequency and time resolution of 
our theory. Eq.~(\ref{eq:reflectivity operator photons}) relates Fourier components 
$\tilde{b}_{\mathrm{in/out}}(\omega)$
that must be understood as averages of the exact Fourier 
components of the incoming and outgoing fields over a bandwidth interval $1/t_{\Delta} \gg 1/t_{\mathrm{fluc}}$. 
For an experiment with a QD linewidth of 1 GHz and a fluctuation time of 1 $\mu$s our theory 
describes effects with a resolution of up to $\sim1$ MHz in frequency and $\sim1$ ns in time.

\section{Optically measured nuclear two-time correlation function}

Having established how frequency components in the incoming and outgoing fields are affected by the nuclear spin bath, we now use this result to show how a measured optical intensity autocorrelation 
depends on a correlation function of the nuclear spins. 
We consider the optical intensity autocorrelation function of the cross-polarised reflected light. 
Assuming a horizontally polarised input field, the correlation in the vertical polarised orientation 
is proportional to the second-order correlation function
\begin{equation}\begin{split}
& G^{(2)}_{V}(t,t+\tau) = \\
& \mathrm{Tr}_{\mathrm{tot}}
\left[ E_V^{(-)}(t) E_V^{(-)}(t+\tau) E_V^{(+)}(t+\tau) E_V^{(+)}(t) \chi \right],
\end{split} \label{eq:G2 general}
\end{equation}
where $E_V^{(\pm)}(t)$ are the positive and negative frequency components of the 
vertically polarised electric field at time $t$, and the trace is performed over the total port mode--cavity--electron spin--nuclear spin 
system, with total initial state $\chi$, and where the Heisenberg electric field operators evolve unitarily in this complete Hilbert space.

To proceed we express these field operators as 
\begin{equation}
E_V^{(+)}(t) = \int_0^{\infty} d \omega \ \tilde{b}^V_{\mathrm{out}}(\omega, t),
 \label{eq:field quad i.t.o. outgoing field operators}
\end{equation}
where we have neglected numerical factors and retardation effects. 
Following Eq.~(\ref{eq:reflectivity operator photons}), a cavity containing a QD with electron 
spin projection $\ket{\uparrow}$ reflects a right circularly polarised photon according to  
$\tilde{b}^R_{\mathrm{out}} (\omega , t) = \hat{r}(\omega, t) \otimes \tilde{b}_{\mathrm{in}}^R$,  
while a left circularly polarised photon acquires a phase shift $r_0(\omega)=\hat{r}(\omega,t)|_{g=0}$ corresponding 
to an empty cavity. Hence we can write 
\begin{equation}
\tilde{b}^V_{\mathrm{out}}(\omega, t) = 
\hat{r}_{\mathrm{cr}}(\omega, t) \otimes \tilde{b}_{\mathrm{in}}^H (\omega) + \hat{r}_{\mathrm{co}}(\omega, t) \otimes \tilde{b}_{\mathrm{in}}^V (\omega),
\label{eq:field quad i.t.o. incoming field operators}
\end{equation}
where the operators $\hat{r}_{\mathrm{co/cr}}(\omega, t) = \frac{1}{2} \left( \hat{r}(\omega ,t) \pm r_0(\omega) \right)$ give 
the reflectivities into the co- and cross-polarised channels, which depend on the nuclear spin state through 
the dependence of $\hat{r}(\omega,t)$ on the Overhauser operator $\smash{\hat{\Delta}(t)}$. 
We assume an initial state 
$\chi = \ket{H(\omega)}\bra{H(\omega)} \otimes \rho_a \otimes \ket{\uparrow}\bra{\uparrow} \otimes \rho_N$,
where $\ket{H(\omega)}$ satisfying  
$\tilde{b}_{\mathrm{in}}^{H}(\omega^{\prime}) \ket{H (\omega)} = \beta \delta (\omega - \omega^{\prime}) \ket{H (\omega)}$ 
is a horizontally polarised coherent state of amplitude $\beta$, $\rho_a$ and $\rho_N$ are states of cavity 
mode and nuclear spin system, respectively, and the electron is assumed to remain in state $\ket{\uparrow}$ 
during the measurement. Substituting this state into Eq.~(\ref{eq:G2 general}) gives 
\begin{align}
\!\!\!& G^{(2)}_{V}(t,t+\tau) = |\beta |^4 \nonumber \\
\!\!\!&  \mathrm{Tr} \!\left[ \hat{r}_{\mathrm{cr}}^{\dagger}(\omega, t) \hat{r}_{\mathrm{cr}}^{\dagger}(\omega, t\!+\!\tau) \hat{r}_{\mathrm{cr}}(\omega, t\!+\!\tau) \hat{r}_{\mathrm{cr}}(\omega, t) \rho_N \right],
\label{eq:G2 i.t.o. nuclear spin exp. val.}
\end{align}
where now and in all that follows the trace is taken only over the nuclear degrees of freedom, showing that we have related an 
optically measured quantity to a nuclear two-time correlation function. 
This correlation function gives the joint probability to measure 
two photons scattered into the cross-polarisation channel at times $t$ and at $t + \tau$. 
It is a {\emph{non-exclusive probability}}, as it does not suppose anything regarding 
any intermediate scattering events, into 
the cross-polarisation channel or otherwise~\cite{plenio_quantum-jump_1998}. Its non-exclusive nature is evidenced 
by the globally unitary evolution of the full Heisenberg picture operator $\smash{E_V^{(-)}(t)}$, 
which depends on the systems involved, including the photonic degrees of freedom. 
A non-exclusive correlation function is the correct form  
to make a connection with experiments, as typically one does not have access to a full scattering history, 
and in practice we take a statistical average over any intermediate scattering events. 

However, we are interested here in how individual scattering events affect the nuclear spin environment, 
which in turn affects later scattering events.  
We therefore seek a relationship between the measured non-exclusive correlation 
function in Eq.~({\ref{eq:G2 i.t.o. nuclear spin exp. val.}}), and an {\emph{exclusive}} correlation function, 
which gives a conditional probability corresponding to a fixed number of scattering events at fixed times~\cite{plenio_quantum-jump_1998}. 
Such a relationship can be expressed as 
\begin{align}
& \!\!\!G^{(2)}_{V}(t,t+\tau) = \sum_{n=0}^{\infty}\sum_{ \{ c_i \} } \nonumber\\
&\!\!\!\!\! \int_t^{t+\tau} \!\!\!\!\!dt_n\! \int_t^{t_n} \!\!\!dt_{n \text{-} 1}...\!\int_t^{t_2} \!\!\!dt_1 \  
 \mathcal{G}^{(n)}_{V,\{ c_i \}, V} (t, t_1,..., t_n, t + \tau),
\label{eq:non-exclusive i.t.o. exclusive}
\end{align}
where $\mathcal{G}^{(n)}_{V,\{ c_i \}, V} (t, t_1, ..., t_n, t + \tau)$ is the exclusive probability 
density that exactly $n+2$ photon scattering events take place in the interval $[t, t+\tau]$, with the first and 
last photons scattered into vertical polarisation at times $t$ and $t+\tau$, 
and $n$ additional photons scattered at intermediate times $t_1,\dots,t_n$ 
into polarisations labelled $\{ c_i\}=c_1,\dots,c_n$ with $c_i$ being either the 
co- ($H$)  or cross-polarised ($V$) channel. 
We now decompose the exclusive probability $\mathcal{G}$ into probabilities describing the scattering times, and the scattering polarisations. 
We write 
\begin{equation}\begin{split}
& \mathcal{G}^{(n)}_{V,\{ c_i \}, V} (t, t_1, ..., t_n, t + \tau) = \\ 
& {p}_t(t,t+\tau;n) \ p_t (t_1, ..., t_n) \ p_c(V, \{ c_i \}, V),
\end{split} \label{eq:exclusive i.t.o. probabilities}
\end{equation}
where $p_t(t,t+\tau;n)$ is the non-exclusive probability of photon scattering events at 
$t$ and $t+\tau$ with $n$ intermediate scattering events at unspecified times, $p_t (t_1, ..., t_n)$ 
the probability density of these intermediate events occurring at $t_1, ..., t_n$, and 
$p_c(V, \{ c_i \}, V)$ the probability of these photons scattered into polarisations $V, c_1,\dots,c_n, V$. 
For a coherent input state $\ket{H (\omega)}$ such as we consider, the probability of 
exactly $n$ scattering events occurring in the interval $[t, t+\tau]$ is given by a 
Poisson distribution $p(n,\tau)$, which depends 
on the coherent state amplitude $\beta$ (related to laser power) and the duration $\tau$, 
while the scattering times are random and 
uncorrelated. This allows us to write ${p}_t(t, t+\tau;n) = G^{(1)}(t)G^{(1)}(t+\tau) p(n,\tau) $ and 
$p_t(t_1, ..., t_n) = n!/\tau^n$, where $G^{(1)}(t)$ is the photon scattering rate. 
The normalised (non-exclusive) cross-polarised intensity correlation function 
can therefore be written 
\begin{align}
&g^{(2)}(\tau) = \frac{G^{(2)}_{V}(t,t+\tau)}{G^{(1)}_V(t)G^{(1)}_V(t+\tau)}=\sum_{n=0}^{\infty} p(n,\tau) \nonumber\\
&\frac{n!}{\tau^n}  
 \int_t^{t+\tau} \!\!\!dt_n \int_t^{t_n} \!\!\!dt_{n \text{-} 1} ... \int_t^{t_2} \!\!\!dt_1 \ \sum_{ \{ c_i \} } \frac{p_c(V, \{ c_i \}, V)}{p_V(t)p_V(t+\tau)},
\label{eq:deg sec order coh}
\end{align}
where $G^{(1)}_V(t) = p_V(t)G^{(1)}(t)$ with $p_V(t)$ the probability that a 
photon scattered at time $t$ is detected in the vertically polarised channel. 
Written in this way, we see that the normalised cross-polarised two-time correlation function 
is the joint probability for two photons to scatter into
the cross-polarisation at times $t$ and $t+\tau$, averaged over all possible numbers, timings, and    
polarisation channels of intermediate events. 

The joint probability $p_c(V, \{ c_i \}, V)$ appearing in Eq.~({\ref{eq:deg sec order coh}}), 
is an exclusive quantity describing the likelihood that exactly $n+2$ photons scatter at times $t, t_1, ..., t_n, t+\tau$ 
with polarisations $V, c_1,\dots,c_n, V$, and can be shown to depend on the nuclear spin system alone. 
To see this, we must understand how the detection of a photon affects the state of the nuclear system, 
and combine this effect with the appropriate nuclear spin evolution in between scattering events. 
In the former case, let us consider the implication 
of the scattering process described in Eq.~({\ref{eq:field quad i.t.o. incoming field operators}}). 
We consider an initially horizontally polarised photon and a nuclear spin state $\ket{\delta}$, giving 
an initial state $\tilde{b}_{\mathrm{in}}^H(\omega)^{\dagger}\ket{0}\otimes \ket{\delta}$, 
with $\ket{0}$ the vacuum. 
If $|\delta\rangle$ is an eigenstate of the Overhauser shift operator $\hat{\Delta}$, we can write  
$\hat{r}_{\mathrm{c}}(\omega)\ket{\delta} = r_{c}^{\delta} \ket{\delta}$ with the subscript indicating 
the co- or crossed-polarised channel. The state after scattering is then given by
\begin{align}
\tilde{b}_{\mathrm{out}}^H(\omega)^{\dagger}\ket{0}\otimes \ket{\delta} =
\Big[r_{\mathrm{cr}}^{\delta} \tilde{b}_{\mathrm{in}}^V(\omega)^{\dagger} + r_{\mathrm{co}}^{\delta} \tilde{b}_{\mathrm{in}}^H(\omega)^{\dagger} \Big] \ket{0} \otimes \ket{\delta}.
\label{eq: scattering process Schrodinger}
\end{align}
From this, we see that destructive (absorptive) detection of a co- or cross-polarised photon from a general nuclear state 
$\ket{\psi}$ then results in an (unnormalized) post-measurement state $\ket{\psi^{\prime}} =  M_{c} \ket{\psi}$ with operators
\begin{equation}
M_{c} = \sum_{\delta} r_{c}^{\delta} \ket{\delta}\bra{\delta}.
 \label{eq:nuclear measurement operator}
\end{equation}
These operators can be interpreted as measurement operators describing 
the effect of a photon scattering event on the nuclear spin system. 
The weights of the associated POVM elements are shown in Fig.~\ref{fig:projector_weight}. 

\begin{figure}
\includegraphics[width=\linewidth]{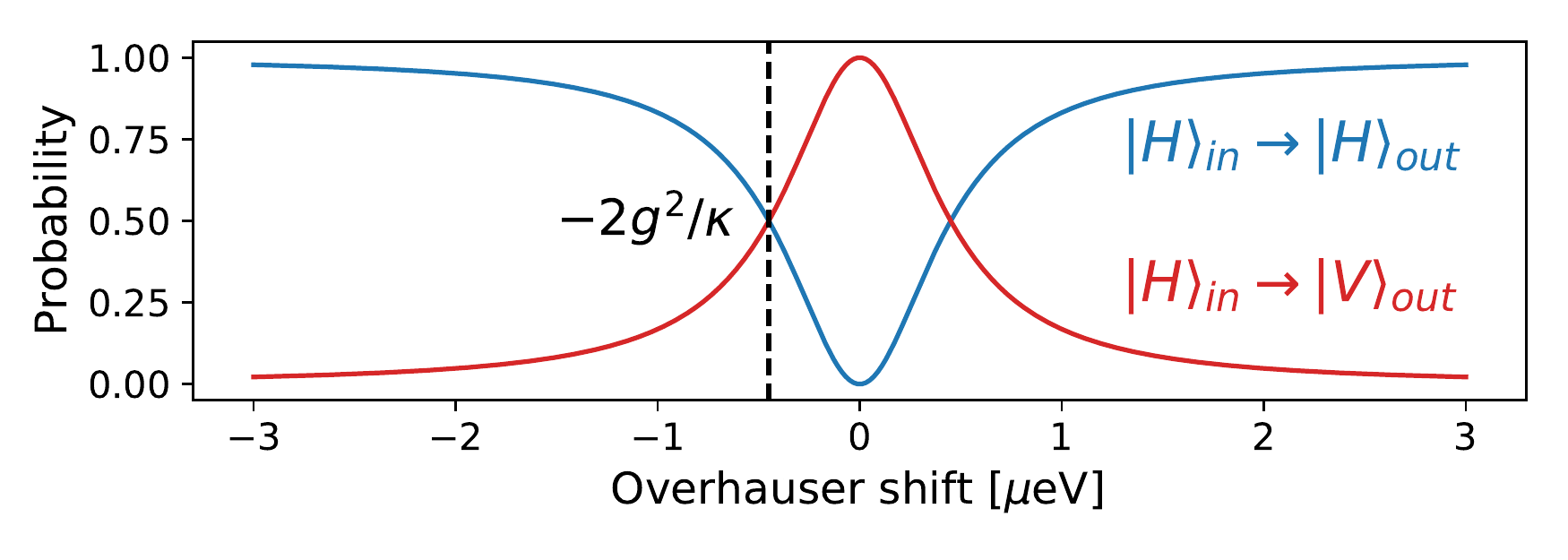}
\caption{Probability for a photon scattering event into the co-polarised $H\to H$ (blue) 
or cross-polarised $H\to V$ channel $\braket{\delta | M^{\dagger}_{c} M_{c} | \delta}$, 
shown for a nuclear system in an Overhauser shift eigenstate $\ket{\delta}$, and 
as a function of the shift $\delta$. 
For zero Overhauser shift cross-polarised photons are preferred, with this bias 
reversing as the Overhauser shift becomes greater than the linewidth of the electronic transition 
in the cavity $2g^2/\kappa$. Parameters [$\mu$eV]: $\kappa = 4000$, $g=30$, $\omega_c = \omega_0 = \omega$.}
\label{fig:projector_weight}
\end{figure}

In between scattering events, since the probability $p(V, \{ c_i \}, V)$ is conditioned on photon scattering events happening 
only at times $t_1,..., t_n$, the nuclear spin evolution is the unitary evolution generated by the Hamiltonian 
$H_N = H_Z + \hat{\Delta}/2 $, where the Zeeman Hamiltonian $H_Z$ and Overhauser shift operator 
$\smash{\hat{\Delta}}$ are defined in Eqs.~(\ref{eq:free and interacting Hamiltonian compononents}) and (\ref{eq:Delta}). 
This allows us to write 
\begin{equation}
p(V, \!\{ c_i \}\!, V) \!=\! \mathrm{Tr} \Big[ \Phi_V\mathcal{U}_{n+1}\Big(\!\prod_{i=1}^n\Phi_{c_i}\mathcal{U}_i\Big)\Phi_V \rho_N (t) \Big],\!
 \label{eq:joint prob evolution operator}
\end{equation}
where the superoperators $\Phi_{c_i}$ and $\mathcal{U}_i$ act as
\begin{equation}
\begin{split}
 \Phi_{c} \rho &= M_{c}\ \rho \ M_{c}^{\dagger}, \\
\mathcal{U}_i \rho &= \e^{- i H_N \Delta_i} \rho\, \e^{ i H_N \Delta_i},
\end{split}
 \label{eq: superoperators}
\end{equation}
and we define $\Delta_i \equiv t_i - t_{i \text{-} 1}$, $t_0 \equiv t$, and $t_{n+1} \equiv t+\tau$. 
The probability in Eq.~({\ref{eq:joint prob evolution operator}}) is exclusive, and corresponds to one possible 
scattering history. The average over all such histories gives the measured two-time correlation 
function following Eq.~(\ref{eq:deg sec order coh}). We note that it is the statistical mixture of 
these histories, and not their coherent superposition, that determines the observed behaviour, 
as for the nuclear spin system the photon scattering events are irreversible measurement processes. 
This formulation is analogous to the quantum jump approach~\cite{plenio_quantum-jump_1998,carmichael_photoelectron_1989}.

\section{Zeno evolution of the nuclear two-time correlation function}

We are now in a position to explore the behaviour of the normalised correlation function $g^{(2)}(\tau)$
given in Eq.~(\ref{eq:deg sec order coh}). We begin by examining the regime of low laser power. 
In this regime we can assume that the probability of intermediate scattering events in a 
time interval $\tau$ vanishes, i.e. $p(0,\tau) \approx 1$ and $p(n \geq 1,\tau) \approx 0$, while the factor involving the product in Eq.~({\ref{eq:joint prob evolution operator}}) is the identity. 
Eq.~(\ref{eq:deg sec order coh}) then gives 
\begin{equation}
g^{(2)}(\tau) \approx 
\frac{p_c(V, \{ \}, V)}{p_V(t)p_V(t+\tau)} = 
\frac{\mathrm{Tr}\left( O_V \mathcal{U}_{0 \rightarrow \tau} \varrho \right)}{\mathrm{Tr}\left( O_V \rho_N \right)^2},
\label{eq: g2 low power}
\end{equation}
where we assume the nuclear system is in a steady state, 
i.e. $\rho(t) = \rho(t+\tau) = \rho_N$, the POVM element is $O_V = M_V^{\dagger} M_V$, 
and we have defined the unnormalised state $\varrho = \Phi_V \rho_N$. 
The steady state assumption allows us to take $t=0$ without loss of generality. Expanding the unitary 
propagator $\mathcal{U}_{0\rightarrow \tau}$ to second order we find 
\begin{align}
\mathrm{Tr}\left( O_V \mathcal{U}_{0\rightarrow \tau}  \varrho \right) \approx\,\,\mathrm{Tr}\left( O_V^2 \rho_N \right) - \frac{1}{2\tau_z^2}\tau^2,
 \label{eq: quadratic short-time evolution}
\end{align} 
where the linear term in $\tau$ vanishes under the assumption that the steady 
state has no coherence in the Overhauser shift eigenbasis, i.e. $[O_V, \rho_N] = 0$, 
and we have defined the nuclear Zeno time
\begin{equation}
\tau_z = 1/\sqrt{\mathrm{Tr}\left( O_V \left[H_N ,[H_N, {\varrho}]\right] \right)}.
\end{equation}
The quadratic short-time behaviour seen in Eq.~({\ref{eq: quadratic short-time evolution}}) 
is characteristic of any unitary evolution, and its experimental 
observation would be a signature of the non-Markovian nature of the nuclear spin bath, and help to 
distinguish it from other sources of resonance fluctuations. Furthermore, 
identification of the timescale of the nuclear spin evolution, given by the Zeno time 
$\tau_z$, would provide valuable information on the dynamical 
behaviour of the nuclear spin system itself.

For laser powers beyond the low intensity regime we need to take intermediate scattering events into account. 
Averaging over the polarisation orientation of the $n$ intermediate events gives 
the polarisation-averaged exclusive probability, which we write as 
\begin{align}
\mathcal{P}_n=\sum_{\{ c_i \}} p(V, \{ c_i \}, V)  
 = \mathrm{Tr} \left( O_V V_{\tau} \varrho \right)
\label{eq:outcome-averaged}
\end{align}
with superoperator $V_{\tau} \equiv \mathcal{U}_{n+1}\prod_{i=1}^n \Phi \ \mathcal{U}_i$. 
Here the superoperator $\Phi$ describes a photon scattering event as a 
non-selective measurement, and acts as 
\begin{equation}
\Phi\ \rho = M_{V} \rho M^{\dagger}_{V} + M_{H} \rho M^{\dagger}_{H}.
 \label{eq:unitary map K}
\end{equation}
Such a non-selective measurement takes a nuclear state $\rho$ and 
rescales all coherences in the Overhauser shift eigenstate basis $\braket{\delta | \rho | \delta^{\prime}}$ by a factor 
\begin{equation}
r_{\delta \delta^{\prime}} = r^{\delta}_{\mathrm{co}} \left( r^{\delta^{\prime}}_{\mathrm{co}}\right)^* 
+ r^{\delta}_{\mathrm{cr}} \left( r^{\delta^{\prime}}_{\mathrm{cr}}\right)^* \equiv |r_{\delta \delta^{\prime}}|\exp (i\theta_{\delta \delta^{\prime}}).
\label{eq:r_ij}
\end{equation}
This factor can be interpreted as an indistinguishability measure 
relating the states $\ket{\delta}$ and $\ket{\delta^{\prime}}$. If both states scatter a photon into the same polarisation, then they cannot be distinguished by photon scattering and $r_{\delta \delta^{\prime}} = 1$. 
On the other hand, if the photons scattered off $\ket{\delta}$ are orthogonal to photons scattered off 
$\ket{\delta^{\prime}}$, then photon scattering has the effect of a projective measurement with 
discarded outcome. In the first case the coherence between $\ket{\delta}$ and $\ket{\delta^{\prime}}$ remains untouched, while 
in the latter the coherence is completely destroyed.

We calculate the probability $\mathcal{P}_n$ to second order in the time intervals $\Delta_i$. 
To do so, we first expand the first time evolution and measurement step to arrive at 
\begin{equation}
V_{\tau}\varrho = \mathcal{U}_{n+1}\prod_{i=2}^n \Phi \ \mathcal{U}_i \left( \varrho + \varrho_1^{(1)} + \varrho_1^{(2)}\right),
 \label{eq:first unitary step expansion}
\end{equation}
where the subscripts indicate the state after the first intermediate photon scattering event, and we have defined the first and second order contributions as 
$ \varrho_1^{(1)} = - i \Delta_1 \Phi \left( [H_N, {\varrho}] \right)$ and $\varrho_1^{(2)}= - (\Delta_1^2/2) \Phi \left( \left[H_N, [H_N, {\varrho}]\right] \right)$. 
Expanding the subsequent steps $\Phi \  \mathcal{U}_i$ and discarding terms of cubic order yields the recursion relations 
\begin{subequations}
\begin{align}
\varrho_k^{(1)} &\!= \Phi \varrho_{k \text{-} 1}^{(1)} - i \Delta_k \Phi \big( [H_N, {\varrho}] \big), \label{eq: rec 1}\\
\varrho_k^{(2)} &\!= \Phi \varrho_{k\text{-}1}^{(2)}\! -\! i \Delta_k \Phi \big( [H_N, \varrho_{k\text{-}1}^{(1)}] \big) - \frac{\Delta_k^2}{2} \Phi \left( \left[H_N, [H_N, {\varrho}]\right] \right). \label{rec 2}
\end{align}
\end{subequations}
In terms of these density operator contributions Eq.~(\ref{eq:outcome-averaged}) becomes 
\begin{equation}\begin{split}
\mathcal{P}_n &= \mathrm{Tr}\left[ O_V \mathcal{U}_{n+1} ( {\varrho} + \varrho_n^{(1)} + \varrho_n^{(2)} ) \right], \\
&= \mathrm{Tr}\left[ O_V {\varrho} \right] + \mathrm{Tr} \left[O_V \varrho_{n+1}^{(2)} \right],
\end{split}
 \label{eq:finally}
\end{equation}
where we make use of the identities 
$\mathrm{Tr}\left[ \Phi A \right] = \mathrm{Tr} \left[ A \right[$, 
$\mathrm{Tr} [ \Phi ( O_V A )] = \mathrm{Tr} [ O_V \Phi (A) ]$ for any operator $A$, and $\mathrm{Tr} [ O_V \rho_k^{(1)} ]= 0$. 
The recursion relations have solutions 
\begin{subequations}
\begin{align}
\varrho_k^{(1)} &= -i \sum_{j=1}^k \Delta_j \Phi^{k\text{-} j+1}\big( [H_N, \varrho] \big), \label{eq:first order dmat}\\
\begin{split}
\varrho_k^{(2)} &= - \sum_{j=1}^k \frac{\Delta_j^2}{2} \Phi^{k \text{-} j+1} \big( \left[H_N, [H_N, {\varrho}]\right] \big) \\
& \ \ - i \sum_{j=1}^k \Delta_j \Phi^{k \text{-} j+1} \big( [H_N, \varrho_{j\text{-}1}^{(1)}] \big), \label{eq:second order dmat}
\end{split}
\end{align}
\end{subequations}
which lead to 
\begin{equation}
\mathcal{P}_n = \mathrm{Tr} \left( O_V V_{t_n}{\varrho} \right) - (\tau - t_n) S_n - \frac{(\tau - t_n)^2}{2 \tau_Z^2},
\label{eq:finally}
\end{equation}
where we define the slope function 
\begin{equation}
S_n \equiv \sum_{j=1}^{n}\Delta_j \mathrm{Tr} \left( O_V \left[ H_N,\Phi^{n\text{-} j+1}[H_N,{\varrho}]\right] \right),
 \label{eq:slope function}
\end{equation}
and where 
\begin{equation}
\mathrm{Tr} \left( O_V V_{t_n}{\varrho} \right) =  \mathrm{Tr} \left( O_V^2 \rho_N \right) -\sum_{k=1}^{n} \Delta_k  S_{k\text{-}1} - \sum_{k=1}^{n}\frac{\Delta_k^2}{2 \tau_Z^2}.
 \label{eq:int1}
\end{equation}

\begin{figure}
\begin{minipage}{0.49 \linewidth}
\includegraphics[width=\linewidth]{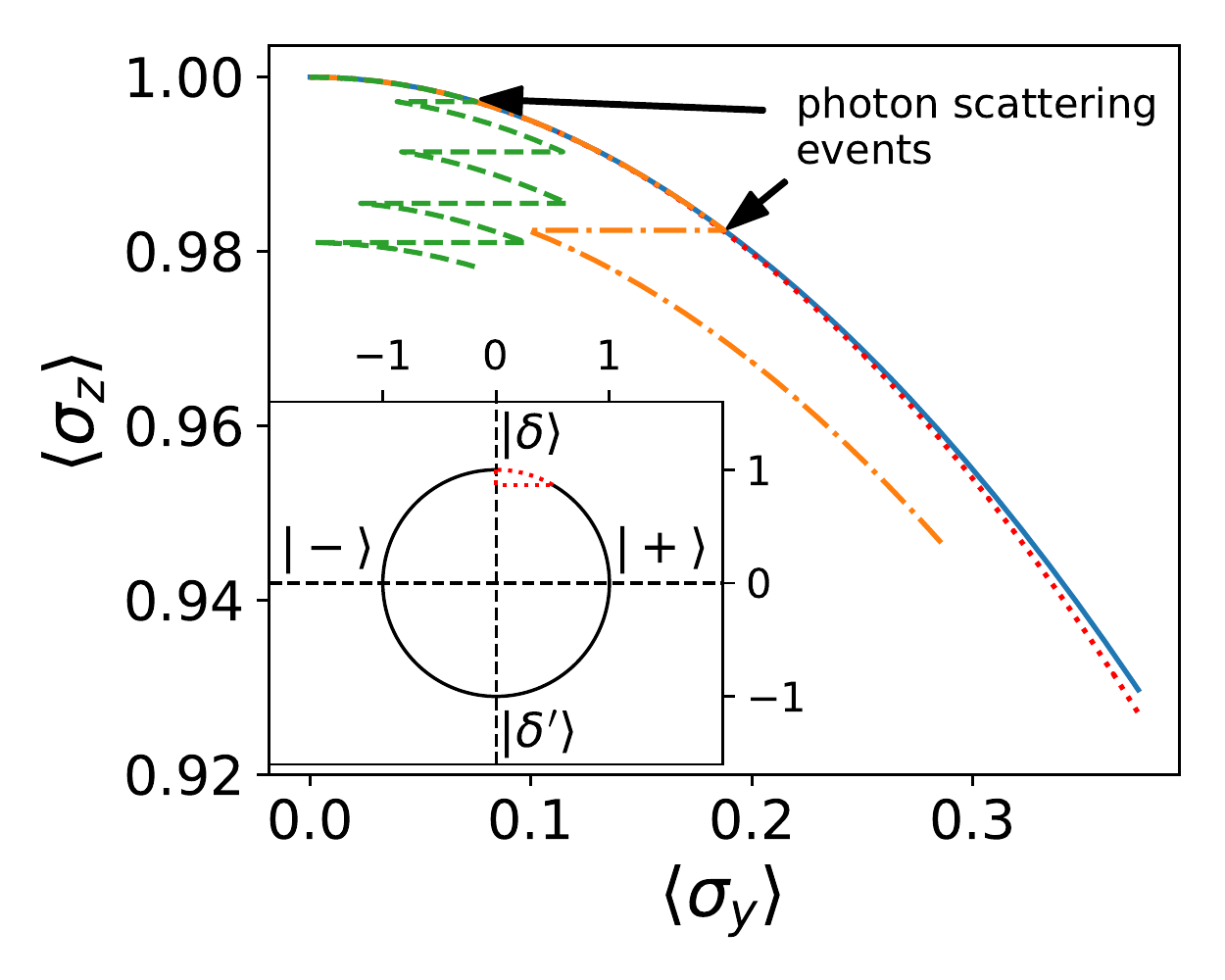}
\end{minipage}
\begin{minipage}{0.49 \linewidth}
\includegraphics[width = \linewidth]{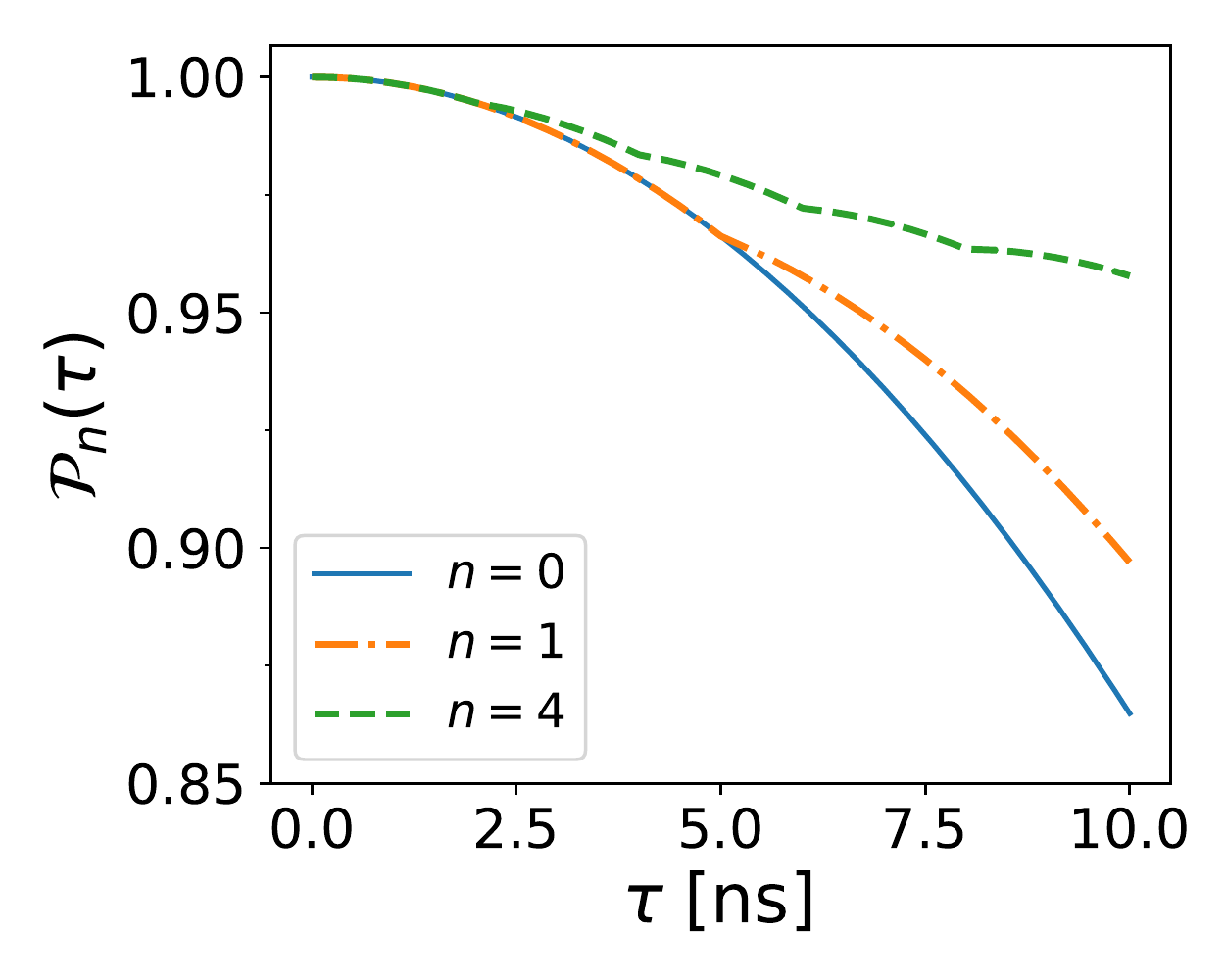}
\end{minipage}
\caption{Left: 
Representative time evolution 
of the nuclear spin system, 
here taken to be two spins spanned 
by the phase shift eigenstates $\ket{\delta}$ and $\ket{\delta^{\prime}}$, 
with the dynamics generated 
following Eqs. \eqref{eq:first order dmat} and \eqref{eq:second order dmat}. 
The plot shows the path of a Bloch vector representation of the nuclear spin state, with 
Overhauser amplitude and coherence operators respectively 
$\sigma_z = \ket{\delta}\bra{\delta} - \ket{\delta^{\prime}}\bra{\delta^{\prime}}$, $\sigma_y = \ket{+}\bra{+} - \ket{-}\bra{-}$, with $\ket{\pm} = \ket{\delta} \pm i \ket{\delta^{\prime}}$. 
State trajectories are shown for no photon scattering events (blue), photon scattering every $\Delta = 5$ ns (orange), and $\Delta = 2$ ns (green). 
The dotted red line shows the surface of the Bloch sphere, a cross-section of which is shown in the inset (red section near the pole corresponds to the outer plot).  Right: Exclusive joint probability $\mathcal{P}_n$ following Eq. \eqref{eq:finally} of scattering into cross-polarisation at times $0$ and $\tau$ corresponding to the state evolution shown to the left, where $\mathcal{P}_n \propto \braket{\sigma_z}$. Parameters: $A_1 = 1$, $A_2 = 3$, $\omega_1=2.5$, $\omega_1=0.5$, $\omega=40$.}
\label{fig:sawtooth}
\end{figure}

\noindent 
Eq.~({\ref{eq:finally}}) constitutes the major result of this work. 
It gives the joint probability of two photons being detected in the vertical (crossed) polarisation channel at times $0$ 
and $\tau$, given $n$ scattering events of unknown polarisation scattering at intermediate times $\{t_i\}$. 
Following Eq.~({\ref{eq:deg sec order coh}}), averaging over the number and timing of these 
intermediate events gives the experimentally measurable cross-polarised intensity autocorrelation function $g_V^{(2)}(\tau)$ shown in Fig. \ref{fig:Zeno}c.

The linear and quadratic terms in Eq.~({\ref{eq:finally}}) can be understood in terms of a generalized 
quantum Zeno effect. As can be seen by the vanishing trace of $\varrho_k^{(1)}$ given in Eq.~({\ref{eq:first order dmat}}), 
the linear order time evolution only affects the coherences of the unnormalised state ${\varrho}$. 
Each measurement $\Phi$ reduces a coherence $\braket{\delta^{\prime}| \varrho_k^{(1)}| \delta}$ by a factor of 
$r_{\delta\delta^{\prime}}$, such that a particular coherence follows a sawtooth pattern shown in Fig. \ref{fig:sawtooth}. The gradient of $\mathcal{P}_n$ at time $\tau$ depends on a commutator of the form $[H_N, \varrho_{\mathrm{coh}}]$, where $\varrho_{\mathrm{coh}}\sim \Phi^{n-j+1}[H_N,\varrho]$ is all the coherence that has accumulated up to $\tau$. 
This coherence has one contribution from the evolution since the last measurement at $t_n$, which leads 
to the quadratic term in Eq.~(\ref{eq:finally}), and another contribution due to all the coherence that has 
partially `survived' the previous measurements, and is given by the linear term. The exponent of 
$\Phi$ gives the number of measurements that the coherence accumulated during interval $\Delta_j$ 
has suffered after the $n$th measurement.

In the limiting case of the polarisation of scattered photons being independent of the nuclear spin state, 
the nuclear spin coherence is not affected by scattering. 
It is readily seen that for $\Phi ([H_N,\rho_N^V]) = [H_N,\rho_N^V]$, 
the slope function becomes $S_n = t_n / \tau_Z^2$ and the quadratic time evolution of 
Eq.~(\ref{eq: quadratic short-time evolution}) is recovered. 
In general, however, an intermediate scattering event and 
associated measurement reduces the coherence, which decreases $S_n$,  
and therefore leads to a reduced slope of $\mathcal{P}_n$. 
This process can be interpreted as the system partially loosing its `memory' of the previous time evolution stored as coherence. 

In the opposite limit, in which projective measurements are made at evenly spaced time intervals $\Delta = \tau/n$, 
coherences are completely destroyed leading to $S_n = 0$, and we find 
\begin{equation}
\mathcal{P}_n \approx \mathrm{Tr} \left( O_v^2 \rho_N \right) - \frac{\Delta}{2 \tau_Z^2} \tau. 
\end{equation}
This linear short-time evolution is characteristic 
of the Markovian regime, in which the system `forgets' all previous time 
evolution with every scattering event. 
The slope of this linear time evolution 
decreases with the number $n$ of measurements, such that frequent photon scattering can 
stabilise the nuclear system in a state that maintains resonance. 
This constitutes a novel nuclear quantum Zeno effect. The results of a Monte-Carlo simulation of 
$g^{(2)}(\tau)$ which averages over the intermediate scattering histories 
are shown in Fig. \ref{fig:Zeno}c, and demonstrate the characteristic flattening of 
the correlation function with increasing laser power, which is the experimental 
signature of this nuclear quantum Zeno effect.

\section{Nuclear quantum Zeno dynamics in the presence of Markovian noise}

The nuclear quantum Zeno dynamics described above arise from unitary evolution of the nuclear spin system 
and the resulting non-Markovian behaviour of the fluctuating excitonic resonance energy. 
However, in typical QD experiments other sources of noise may be present which lead to dephasing of the excition, and which are Markovian and memoryless on the timescale of the nuclear spin evolution. In particular fluctuating charges in the vicinity of the QD can dephase the excitonic state~\cite{kuhlmann_transform-limited_2015}, and also phonons can perturb the excitonic transition \cite{nazir_modelling_2016}.
To investigate the effects of Markovian dephasing noise, we add a random, time varying shift $s(t)$ to the resonance energy $\omega_0$ that takes on a particular value with probability $p(s)$. This shift leads to dephasing of the excitonic state, as the phase of that state evolves in proportion to the exciton energy. Given the random shift of this energy, the phase undergoes a random walk and the average state dephases at a rate $\gamma = t_c \sigma^2 /2$, where $\sigma^2$ is the variance of the shift probability distribution $p(s)$ and $t_c$ is the characteristic timescale of the fluctuations. In our case it is the typical magnitude $\sigma$ of the shift rather than the dephasing rate $\gamma$ which affects the nuclear spin evolution, as explained below.

Given a nuclear spin state $\ket{\delta}$ and a random energy shift $s$, a horizontally polarised photon $\ket{H}$ is scattered into a polarisation state $r_{\mathrm{cr}}^{\delta + s} \ket{V} + r_{\mathrm{co}}^{\delta + s} \ket{H}$ with probability $p(s)$, where $r_{\mathrm{co / cr}}^{\delta + s}$ are the reflection coefficients into co/cross-polarisation, respectively (cf. Eq. \ref{eq: scattering process Schrodinger}). 
The probability $p_{\mathrm{cr}}$ of photon scattering into cross-polarisation is then associated with a modified 
POVM element $\tilde{O}_V = \sum_{\delta}\sum_s p(s) |r_{\mathrm{cr}}^{\delta + s}|^2 \ket{\delta} \bra{\delta}$ and the nuclear spin state upon scattering is obtained by the quantum operation $\tilde{\Phi}_V$ with operators $\{ M_V^s = \sqrt{p(s)} \sum_{\delta} r_{cr}^{\delta+s} \ket{\delta}\bra{\delta} \}$, i.e.
\begin{equation}
\rho^{\prime}_{\mathrm{cr}} = \frac{1}{p_{\mathrm{cr}}} \tilde{\Phi}_V \rho =  \frac{1}{\mathrm{Tr}\left( \tilde{O}_V \rho \right)} \sum_s M_V^s \rho \left( M_V^s \right)^{\dagger}
 \label{eq: post mm state stochastic shift}
\end{equation}
(cf. Eq. \ref{eq:nuclear measurement operator} and \ref{eq: g2 low power}). The random resonance shift $s$ effectively broadens the resonant feature in the polarization rotation.  Replacing the POVM element $O_V$ and quantum operation $\Phi$ in Eq. \ref{eq:outcome-averaged} by their stochastic versions $\tilde{O}_V$ and $\tilde{\Phi}$ then 
yields a modified cross-polarized intensity autocorrelation $\smash{g_V^{(2)}(\tau)}$. 

\begin{figure}
\includegraphics[width=0.8\linewidth]{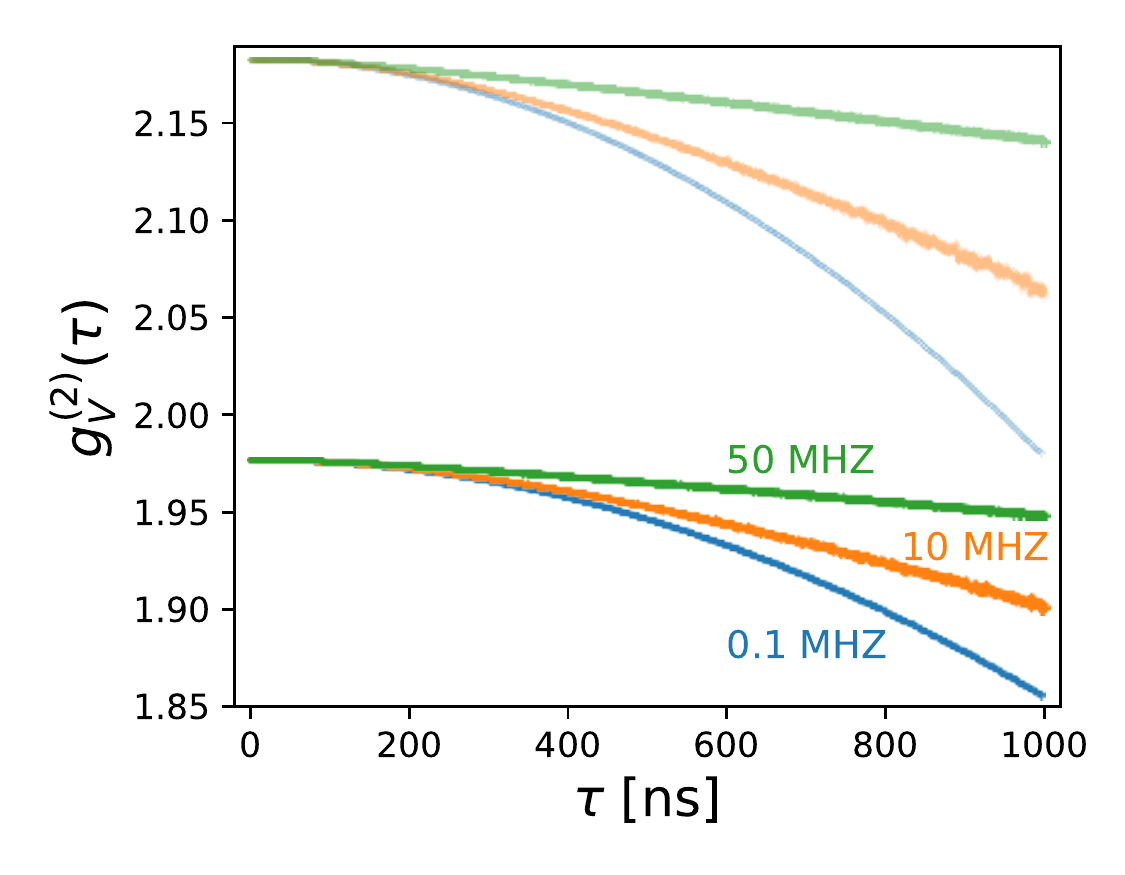}
\caption{Cross-polarised optical intensity autocorrelation $g_V^{(2)}(\tau)$ including Markovian resonance fluctuations noise. 
The transparent lines show the data from Fig. \ref{fig:Zeno}c) without Markovian noise for comparison. 
A random, uncorrelated resonance shift $s$ is sampled from from a Gaussian distribution with zero mean and 
$250$ MHz variance at each photon scattering event of the Monte-Carlo simulation, with all other parameters as in Fig. \ref{fig:Zeno}c).}
 \label{fig:g2 with stochastic noise}
\end{figure}

The result of a Monte Carlo simulation including these Markovian processes is shown in Fig. \ref{fig:g2 with stochastic noise}, 
and we see that they have a twofold effect on the intensity autocorrelation. Firstly, the bunching seen at 
$\tau=0$ is reduced, which is a consequence of the Markovianity of the noise on the timescale of the nuclear spin evolution. 
A cross-polarised photon detection projects the nuclear spin system into a state with significantly increased probability weight on resonant configurations, 
such that a photon scattering event immediately afterwards has a high chance of scattering into the cross-polarised channel. 
In the presence of stochastic noise, however, a cross-polarised photon detection yields less information regarding the nuclear spin state, 
and therefore leads to a smaller increase in the likelihood of a second cross-polarised scattering event, which results in weaker bunching and 
thus a decreased $\smash{g^{(2)}_V(0)}$. Secondly, the effect of intermediate photon scattering events impeding the nuclear spin evolution, and thereby the decay of $g^{(2)}_V(\tau)$ via the quantum Zeno effect, is reduced. This reduced effect of a photon scattering event on the nuclear spin evolution can be attributed to the 
decreased precision of the measurement performed on the nuclear spin system by the photon due to the averaging over stochastic shifts $s$. 
As can be seen in Fig.~\ref{fig:g2 with stochastic noise}, however, neither of these effects 
alter the qualitative behaviour of the correlation function. The decay of the intensity autocorrelation still 
changes from relatively fast and initially quadratic for low laser intensities, to slow and exponential for higher intensities. 
Observation of this quantum Zeno effect should therefore be possible if the broadening of the excitionic transition due to dephasing is significantly 
smaller than the broadening due to nuclear spins, such that a photon scattering event still yields sufficient 
information of the nuclear spin state.

The time evolution of $g^{(2)}_V(\tau)$ in the above analysis is due to nuclear spin evolution alone, and the electron spin is assumed 
to remain in an eigenstate of the electron Zeeman Hamiltonian. In an isolated electron--nuclear spin system with a magnetic field 
$B_{\mathrm{ext}} \gtrsim 100$ mT \cite{cywinski_pure_2009}, this assumption is well justified, 
as electron spin relaxation by electron--nuclear spin flip flops is energetically forbidden. In practice, 
however, there are additional mechanisms that 
may lead to electron spin relaxation, which in turn will lead to an exponentially decaying intensity correlation 
function that cannot be stabilized by the nuclear quantum Zeno effect. 
For example, co-tunneling of electrons in and out of the QD can lead to such electron spin relaxation, 
although we note that this mechanism can be strongly suppressed by tuning of the QD energy with an external electric field \cite{dreiser_optical_2008}.

Another electron spin relaxation mechanism that might obscure the nuclear quantum Zeno dynamics is given by second-order electron-nuclear flip flops, which arise when an environment such as the phonon bath supplies or absorbs the flip flop energy~\cite{latta_confluence_2009}. 
Such environment-assisted flip flops arise from the contact hyperfine Hamiltonian given by
\begin{equation}
H_{hf} = \sum_k A_k \left( S^z I_k^z + \frac{1}{2} \left( S^- I_k^+ + S^+ I_k^- \right) \right),
 \label{eq:contact hyperfine Hamiltonian}
\end{equation}
when also in the presence of electron spin dephasing at a rate $\eta$, 
and lead to electron spin relaxation at a rate $\sim \eta \frac{\alpha^2}{\omega_e^2}$, where $\omega_e$ is the electronic Zeeman splitting as before 
and $\alpha$ gives the interaction energy with the unpolarised nuclear spin bath \cite{nutz_solvable_2019}. 
This relaxation mechanism is therefore suppressed by a strong external magnetic field and tuning of the temperature and electrostatic environment to minimise the dephasing rate, 
however a theoretical estimate of $\eta$ and $\alpha$ is beyond the scope of this work. 
Experimental measurement of the electron spin relaxation rate has confirmed the $\omega_e^{-2}$ suppression of relaxation by a magnetic field and achieved spin lifetimes of hundreds of $\mu$s  at sub-Tesla magnetic fields \cite{dreiser_optical_2008}, which is significantly slower than the resonance fluctuations 
that are important in this work~\cite{androvitsaneas_efficient_2019}. 
Putting these observations together, we conclude that the nuclear quantum Zeno effect should therefore be observable for magnetic fields of 
$\sim 1$ T, temperatures of $T=4~\mathrm{K}$, and using tuning of the charge state to maximise the electron spin lifetime.

\section{Discussion and Conclusion}

In order to experimentally demonstrate the nuclear quantum Zeno effect predicted here, 
it would suffice to observe the characteristic change of the cross polarised intensity autocorrelation function from quadratic to linear short-time behaviour, as the intensity of the input laser light increases. 
The non-Markovian quadratic regime is the most challenging to observe, since the intensity must be low enough that unobserved intermediate scattering events have vanishing probability, which in turn implies a long integration time of the experiment.
Increasing the input laser intensity will introduce intermediate photon scattering events that take place 
during a delay time $\tau$ of interest. If these photons can be detected, the polarisation outcomes of these 
detection events need to be averaged over to calculate the degree of second-order coherence 
$g^{(2)}(\tau)$. If these events are lost and not detected, this averaging is automatically performed. 
Loss does therefore not invalidate the measurement as long as there is an estimate of the photon scattering rate for a given intensity. Following Eq.~(\ref{eq:finally}) one 
expects a broadening as well as a change from quadratic to linear behaviour of the intensity 
autocorrelation function with intensity, which is a second experimental feature of the nuclear quantum Zeno effect.

Our result paves the way for experimental demonstration of a novel nuclear spin effect in quantum dots, 
with implications for both fundamental 
theoretical investigations and photonic quantum computing. 
Importantly, our formulation of the quantum Zeno effect in terms of a two-time correlation function has the 
advantage that it is possible to observe the effect without initialising the system in a particular state. The intensity 
autocorrelation considers pairs of cross-polarised photon detection events, the first of which effectively 
initializes the nuclear system in a state with increased likelihood of being close to resonance. The second 
photon count then probes how far the system has evolved away from this initial state, and intermediate 
photon counts disturb this evolution. This generalized description of a quantum Zeno effect in terms of 
imperfect measurements and two-time correlation functions likely applies to other experimentally 
accessible quantum systems. Another interesting theoretical aspect of the nuclear quantum Zeno 
process is the explicit connection between a measurement and the physical process of photon scattering. 
This connection shows that it is the coherence-destroying effect of measurements that impedes 
coherent evolution and gives rise to the quantum Zeno effect. The formulation of coherence reduction 
of photon scattering as a measurement is merely a convenient formalism, making it clear 
that a coherence-removing process that gives rise to a quantum 
Zeno effect does not need to be a measurement.

Beyond these implications 
the nuclear quantum Zeno effect may be relevant 
to the experimental realization of a quantum dot-based source of entangled photons. 
A weak laser could be used to stabilise 
the nuclear system in a state for which the electronic transition is close to resonance and where high phase 
shifts can be achieved. If a method was found to simultaneously keep the electron spin in a superposition 
then the nuclear Zeno effect could be used to realize photonic states with useful entanglement properties 
as proposed in~\cite{hu_deterministic_2008}, even in the presence of a nuclear spin environment.

\begin{acknowledgements}
The authors thank John Rarity, Terry Rudolph, Sophia Economou, Ed Barnes, Will McCutcheon, and Gary Sinclair for interesting and useful discussions. 
T.N. acknowledges financial support from the People Programme (Marie Curie Actions) of the European Unions
Seventh Framework Programme (FP7/2007-2013) under
REA Grant agreement 317232. This project has received funding from the 
European Union's Horizon 2020 research and innovation programme under the 
Marie Sk{\l}odowska-Curie grant agreement No. 703193. This work was also 
funded by the Frontiers in Quantum Technologies programme and by the Engineering and Physical Sciences Research Council (EP/M024156/1, EP/N003381/1 and EP/L024020/1).
\end{acknowledgements}

\bibliography{Zeno}

\begin{thebibliography}{41}%
\makeatletter
\providecommand \@ifxundefined [1]{%
 \@ifx{#1\undefined}
}%
\providecommand \@ifnum [1]{%
 \ifnum #1\expandafter \@firstoftwo
 \else \expandafter \@secondoftwo
 \fi
}%
\providecommand \@ifx [1]{%
 \ifx #1\expandafter \@firstoftwo
 \else \expandafter \@secondoftwo
 \fi
}%
\providecommand \natexlab [1]{#1}%
\providecommand \enquote  [1]{``#1''}%
\providecommand \bibnamefont  [1]{#1}%
\providecommand \bibfnamefont [1]{#1}%
\providecommand \citenamefont [1]{#1}%
\providecommand \href@noop [0]{\@secondoftwo}%
\providecommand \href [0]{\begingroup \@sanitize@url \@href}%
\providecommand \@href[1]{\@@startlink{#1}\@@href}%
\providecommand \@@href[1]{\endgroup#1\@@endlink}%
\providecommand \@sanitize@url [0]{\catcode `\\12\catcode `\$12\catcode
  `\&12\catcode `\#12\catcode `\^12\catcode `\_12\catcode `\%12\relax}%
\providecommand \@@startlink[1]{}%
\providecommand \@@endlink[0]{}%
\providecommand \url  [0]{\begingroup\@sanitize@url \@url }%
\providecommand \@url [1]{\endgroup\@href {#1}{\urlprefix }}%
\providecommand \urlprefix  [0]{URL }%
\providecommand \Eprint [0]{\href }%
\providecommand \doibase [0]{http://dx.doi.org/}%
\providecommand \selectlanguage [0]{\@gobble}%
\providecommand \bibinfo  [0]{\@secondoftwo}%
\providecommand \bibfield  [0]{\@secondoftwo}%
\providecommand \translation [1]{[#1]}%
\providecommand \BibitemOpen [0]{}%
\providecommand \bibitemStop [0]{}%
\providecommand \bibitemNoStop [0]{.\EOS\space}%
\providecommand \EOS [0]{\spacefactor3000\relax}%
\providecommand \BibitemShut  [1]{\csname bibitem#1\endcsname}%
\let\auto@bib@innerbib\@empty
\bibitem [{\citenamefont {Lindner}\ and\ \citenamefont
  {Rudolph}(2009)}]{lindner_proposal_2009}%
  \BibitemOpen
  \bibfield  {author} {\bibinfo {author} {\bibfnamefont {N.~H.}\ \bibnamefont
  {Lindner}}\ and\ \bibinfo {author} {\bibfnamefont {T.}~\bibnamefont
  {Rudolph}},\ }\href {\doibase 10.1103/PhysRevLett.103.113602} {\bibfield
  {journal} {\bibinfo  {journal} {Physical Review Letters}\ }\textbf {\bibinfo
  {volume} {103}},\ \bibinfo {pages} {113602} (\bibinfo {year}
  {2009})}\BibitemShut {NoStop}%
\bibitem [{\citenamefont {Economou}\ \emph {et~al.}(2010)\citenamefont
  {Economou}, \citenamefont {Lindner},\ and\ \citenamefont
  {Rudolph}}]{economou_optically_2010}%
  \BibitemOpen
  \bibfield  {author} {\bibinfo {author} {\bibfnamefont {S.~E.}\ \bibnamefont
  {Economou}}, \bibinfo {author} {\bibfnamefont {N.}~\bibnamefont {Lindner}}, \
  and\ \bibinfo {author} {\bibfnamefont {T.}~\bibnamefont {Rudolph}},\ }\href
  {\doibase 10.1103/PhysRevLett.105.093601} {\bibfield  {journal} {\bibinfo
  {journal} {Physical Review Letters}\ }\textbf {\bibinfo {volume} {105}},\
  \bibinfo {pages} {093601} (\bibinfo {year} {2010})}\BibitemShut {NoStop}%
\bibitem [{\citenamefont {Schwartz}\ \emph {et~al.}(2016)\citenamefont
  {Schwartz}, \citenamefont {Cogan}, \citenamefont {Schmidgall}, \citenamefont
  {Don}, \citenamefont {Gantz}, \citenamefont {Kenneth}, \citenamefont
  {Lindner},\ and\ \citenamefont {Gershoni}}]{schwartz_deterministic_2016-2}%
  \BibitemOpen
  \bibfield  {author} {\bibinfo {author} {\bibfnamefont {I.}~\bibnamefont
  {Schwartz}}, \bibinfo {author} {\bibfnamefont {D.}~\bibnamefont {Cogan}},
  \bibinfo {author} {\bibfnamefont {E.~R.}\ \bibnamefont {Schmidgall}},
  \bibinfo {author} {\bibfnamefont {Y.}~\bibnamefont {Don}}, \bibinfo {author}
  {\bibfnamefont {L.}~\bibnamefont {Gantz}}, \bibinfo {author} {\bibfnamefont
  {O.}~\bibnamefont {Kenneth}}, \bibinfo {author} {\bibfnamefont {N.~H.}\
  \bibnamefont {Lindner}}, \ and\ \bibinfo {author} {\bibfnamefont
  {D.}~\bibnamefont {Gershoni}},\ }\href {\doibase 10.1126/science.aah4758}
  {\bibfield  {journal} {\bibinfo  {journal} {Science}\ }\textbf {\bibinfo
  {volume} {354}},\ \bibinfo {pages} {434} (\bibinfo {year}
  {2016})}\BibitemShut {NoStop}%
\bibitem [{\citenamefont {Hu}\ \emph {et~al.}(2008{\natexlab{a}})\citenamefont
  {Hu}, \citenamefont {Young}, \citenamefont {O’Brien}, \citenamefont
  {Munro},\ and\ \citenamefont {Rarity}}]{hu_giant_2008}%
  \BibitemOpen
  \bibfield  {author} {\bibinfo {author} {\bibfnamefont {C.~Y.}\ \bibnamefont
  {Hu}}, \bibinfo {author} {\bibfnamefont {A.}~\bibnamefont {Young}}, \bibinfo
  {author} {\bibfnamefont {J.~L.}\ \bibnamefont {O’Brien}}, \bibinfo {author}
  {\bibfnamefont {W.~J.}\ \bibnamefont {Munro}}, \ and\ \bibinfo {author}
  {\bibfnamefont {J.~G.}\ \bibnamefont {Rarity}},\ }\href {\doibase
  10.1103/PhysRevB.78.085307} {\bibfield  {journal} {\bibinfo  {journal}
  {Physical Review B}\ }\textbf {\bibinfo {volume} {78}},\ \bibinfo {pages}
  {085307} (\bibinfo {year} {2008}{\natexlab{a}})}\BibitemShut {NoStop}%
\bibitem [{\citenamefont {Hu}\ \emph {et~al.}(2008{\natexlab{b}})\citenamefont
  {Hu}, \citenamefont {Munro},\ and\ \citenamefont
  {Rarity}}]{hu_deterministic_2008}%
  \BibitemOpen
  \bibfield  {author} {\bibinfo {author} {\bibfnamefont {C.~Y.}\ \bibnamefont
  {Hu}}, \bibinfo {author} {\bibfnamefont {W.~J.}\ \bibnamefont {Munro}}, \
  and\ \bibinfo {author} {\bibfnamefont {J.~G.}\ \bibnamefont {Rarity}},\
  }\href {\doibase 10.1103/PhysRevB.78.125318} {\bibfield  {journal} {\bibinfo
  {journal} {Physical Review B}\ }\textbf {\bibinfo {volume} {78}},\ \bibinfo
  {pages} {125318} (\bibinfo {year} {2008}{\natexlab{b}})}\BibitemShut
  {NoStop}%
\bibitem [{\citenamefont {Pineiro-Orioli}\ \emph {et~al.}(2013)\citenamefont
  {Pineiro-Orioli}, \citenamefont {McCutcheon},\ and\ \citenamefont
  {Rudolph}}]{pineiro-orioli_noise_2013}%
  \BibitemOpen
  \bibfield  {author} {\bibinfo {author} {\bibfnamefont {A.}~\bibnamefont
  {Pineiro-Orioli}}, \bibinfo {author} {\bibfnamefont {D.~P.~S.}\ \bibnamefont
  {McCutcheon}}, \ and\ \bibinfo {author} {\bibfnamefont {T.}~\bibnamefont
  {Rudolph}},\ }\href {\doibase 10.1103/PhysRevB.88.035315} {\bibfield
  {journal} {\bibinfo  {journal} {Physical Review B}\ }\textbf {\bibinfo
  {volume} {88}},\ \bibinfo {pages} {035315} (\bibinfo {year}
  {2013})}\BibitemShut {NoStop}%
\bibitem [{\citenamefont {Lodahl}\ \emph {et~al.}(2015)\citenamefont {Lodahl},
  \citenamefont {Mahmoodian},\ and\ \citenamefont
  {Stobbe}}]{lodahl_interfacing_2015}%
  \BibitemOpen
  \bibfield  {author} {\bibinfo {author} {\bibfnamefont {P.}~\bibnamefont
  {Lodahl}}, \bibinfo {author} {\bibfnamefont {S.}~\bibnamefont {Mahmoodian}},
  \ and\ \bibinfo {author} {\bibfnamefont {S.}~\bibnamefont {Stobbe}},\ }\href
  {\doibase 10.1103/RevModPhys.87.347} {\bibfield  {journal} {\bibinfo
  {journal} {Reviews of Modern Physics}\ }\textbf {\bibinfo {volume} {87}},\
  \bibinfo {pages} {347} (\bibinfo {year} {2015})}\BibitemShut {NoStop}%
\bibitem [{\citenamefont {Lodahl}(2018)}]{lodahl_quantum-dot_2018}%
  \BibitemOpen
  \bibfield  {author} {\bibinfo {author} {\bibfnamefont {P.}~\bibnamefont
  {Lodahl}},\ }\href {\doibase 10.1088/2058-9565/aa91bb} {\bibfield  {journal}
  {\bibinfo  {journal} {Quantum Science and Technology}\ }\textbf {\bibinfo
  {volume} {3}},\ \bibinfo {pages} {013001} (\bibinfo {year}
  {2018})}\BibitemShut {NoStop}%
\bibitem [{\citenamefont {McCutcheon}\ \emph {et~al.}(2014)\citenamefont
  {McCutcheon}, \citenamefont {Lindner},\ and\ \citenamefont
  {Rudolph}}]{mccutcheon_error_2014}%
  \BibitemOpen
  \bibfield  {author} {\bibinfo {author} {\bibfnamefont {D.~P.~S.}\
  \bibnamefont {McCutcheon}}, \bibinfo {author} {\bibfnamefont {N.~H.}\
  \bibnamefont {Lindner}}, \ and\ \bibinfo {author} {\bibfnamefont
  {T.}~\bibnamefont {Rudolph}},\ }\href {\doibase
  10.1103/PhysRevLett.113.260503} {\bibfield  {journal} {\bibinfo  {journal}
  {Physical Review Letters}\ }\textbf {\bibinfo {volume} {113}},\ \bibinfo
  {pages} {260503} (\bibinfo {year} {2014})}\BibitemShut {NoStop}%
\bibitem [{\citenamefont {Kuhlmann}\ \emph {et~al.}(2015)\citenamefont
  {Kuhlmann}, \citenamefont {Prechtel}, \citenamefont {Houel}, \citenamefont
  {Ludwig}, \citenamefont {Reuter}, \citenamefont {Wieck},\ and\ \citenamefont
  {Warburton}}]{kuhlmann_transform-limited_2015}%
  \BibitemOpen
  \bibfield  {author} {\bibinfo {author} {\bibfnamefont {A.~V.}\ \bibnamefont
  {Kuhlmann}}, \bibinfo {author} {\bibfnamefont {J.~H.}\ \bibnamefont
  {Prechtel}}, \bibinfo {author} {\bibfnamefont {J.}~\bibnamefont {Houel}},
  \bibinfo {author} {\bibfnamefont {A.}~\bibnamefont {Ludwig}}, \bibinfo
  {author} {\bibfnamefont {D.}~\bibnamefont {Reuter}}, \bibinfo {author}
  {\bibfnamefont {A.~D.}\ \bibnamefont {Wieck}}, \ and\ \bibinfo {author}
  {\bibfnamefont {R.~J.}\ \bibnamefont {Warburton}},\ }\href {\doibase
  10.1038/ncomms9204} {\bibfield  {journal} {\bibinfo  {journal} {Nature
  Communications}\ }\textbf {\bibinfo {volume} {6}},\ \bibinfo {pages} {8204}
  (\bibinfo {year} {2015})}\BibitemShut {NoStop}%
\bibitem [{\citenamefont {Stockill}\ \emph {et~al.}(2016)\citenamefont
  {Stockill}, \citenamefont {Le~Gall}, \citenamefont {Matthiesen},
  \citenamefont {Huthmacher}, \citenamefont {Clarke}, \citenamefont {Hugues},\
  and\ \citenamefont {Atat{\"u}re}}]{stockill_quantum_2016}%
  \BibitemOpen
  \bibfield  {author} {\bibinfo {author} {\bibfnamefont {R.}~\bibnamefont
  {Stockill}}, \bibinfo {author} {\bibfnamefont {C.}~\bibnamefont {Le~Gall}},
  \bibinfo {author} {\bibfnamefont {C.}~\bibnamefont {Matthiesen}}, \bibinfo
  {author} {\bibfnamefont {L.}~\bibnamefont {Huthmacher}}, \bibinfo {author}
  {\bibfnamefont {E.}~\bibnamefont {Clarke}}, \bibinfo {author} {\bibfnamefont
  {M.}~\bibnamefont {Hugues}}, \ and\ \bibinfo {author} {\bibfnamefont
  {M.}~\bibnamefont {Atat{\"u}re}},\ }\href {\doibase 10.1038/ncomms12745}
  {\bibfield  {journal} {\bibinfo  {journal} {Nature Communications}\ }\textbf
  {\bibinfo {volume} {7}},\ \bibinfo {pages} {12745} (\bibinfo {year}
  {2016})}\BibitemShut {NoStop}%
\bibitem [{\citenamefont {Wuest}\ \emph {et~al.}(2016)\citenamefont {Wuest},
  \citenamefont {Munsch}, \citenamefont {Maier}, \citenamefont {Kuhlmann},
  \citenamefont {Ludwig}, \citenamefont {Wieck}, \citenamefont {Loss},
  \citenamefont {Poggio},\ and\ \citenamefont {Warburton}}]{wuest_role_2016}%
  \BibitemOpen
  \bibfield  {author} {\bibinfo {author} {\bibfnamefont {G.}~\bibnamefont
  {Wuest}}, \bibinfo {author} {\bibfnamefont {M.}~\bibnamefont {Munsch}},
  \bibinfo {author} {\bibfnamefont {F.}~\bibnamefont {Maier}}, \bibinfo
  {author} {\bibfnamefont {A.~V.}\ \bibnamefont {Kuhlmann}}, \bibinfo {author}
  {\bibfnamefont {A.}~\bibnamefont {Ludwig}}, \bibinfo {author} {\bibfnamefont
  {A.~D.}\ \bibnamefont {Wieck}}, \bibinfo {author} {\bibfnamefont
  {D.}~\bibnamefont {Loss}}, \bibinfo {author} {\bibfnamefont {M.}~\bibnamefont
  {Poggio}}, \ and\ \bibinfo {author} {\bibfnamefont {R.~J.}\ \bibnamefont
  {Warburton}},\ }\href {\doibase 10.1038/nnano.2016.114} {\bibfield  {journal}
  {\bibinfo  {journal} {Nature Nanotechnology}\ }\textbf {\bibinfo {volume}
  {11}},\ \bibinfo {pages} {885} (\bibinfo {year} {2016})}\BibitemShut
  {NoStop}%
\bibitem [{\citenamefont {Ethier-Majcher}\ \emph {et~al.}(2017)\citenamefont
  {Ethier-Majcher}, \citenamefont {Gangloff}, \citenamefont {Stockill},
  \citenamefont {Clarke}, \citenamefont {Hugues}, \citenamefont {Le~Gall},\
  and\ \citenamefont {Atat{\"u}re}}]{ethier-majcher_improving_2017}%
  \BibitemOpen
  \bibfield  {author} {\bibinfo {author} {\bibfnamefont {G.}~\bibnamefont
  {Ethier-Majcher}}, \bibinfo {author} {\bibfnamefont {D.}~\bibnamefont
  {Gangloff}}, \bibinfo {author} {\bibfnamefont {R.}~\bibnamefont {Stockill}},
  \bibinfo {author} {\bibfnamefont {E.}~\bibnamefont {Clarke}}, \bibinfo
  {author} {\bibfnamefont {M.}~\bibnamefont {Hugues}}, \bibinfo {author}
  {\bibfnamefont {C.}~\bibnamefont {Le~Gall}}, \ and\ \bibinfo {author}
  {\bibfnamefont {M.}~\bibnamefont {Atat{\"u}re}},\ }\href {\doibase
  10.1103/PhysRevLett.119.130503} {\bibfield  {journal} {\bibinfo  {journal}
  {Physical Review Letters}\ }\textbf {\bibinfo {volume} {119}},\ \bibinfo
  {pages} {130503} (\bibinfo {year} {2017})}\BibitemShut {NoStop}%
\bibitem [{\citenamefont {Greilich}\ \emph {et~al.}(2007)\citenamefont
  {Greilich}, \citenamefont {Silva}, \citenamefont {Moussa}, \citenamefont
  {Ryan}, \citenamefont {Laforest}, \citenamefont {Baugh}, \citenamefont
  {Cory},\ and\ \citenamefont {Laflamme}}]{greilich_nuclei-induced_2007}%
  \BibitemOpen
  \bibfield  {author} {\bibinfo {author} {\bibfnamefont {J.}~\bibnamefont
  {Greilich}}, \bibinfo {author} {\bibfnamefont {M.}~\bibnamefont {Silva}},
  \bibinfo {author} {\bibfnamefont {O.}~\bibnamefont {Moussa}}, \bibinfo
  {author} {\bibfnamefont {C.}~\bibnamefont {Ryan}}, \bibinfo {author}
  {\bibfnamefont {M.}~\bibnamefont {Laforest}}, \bibinfo {author}
  {\bibfnamefont {J.}~\bibnamefont {Baugh}}, \bibinfo {author} {\bibfnamefont
  {D.~G.}\ \bibnamefont {Cory}}, \ and\ \bibinfo {author} {\bibfnamefont
  {R.}~\bibnamefont {Laflamme}},\ }\href {\doibase 10.1126/science.1145699}
  {\bibfield  {journal} {\bibinfo  {journal} {Science}\ }\textbf {\bibinfo
  {volume} {317}},\ \bibinfo {pages} {1893} (\bibinfo {year}
  {2007})}\BibitemShut {NoStop}%
\bibitem [{\citenamefont {Barnes}\ and\ \citenamefont
  {Economou}(2011)}]{barnes_electron-nuclear_2011}%
  \BibitemOpen
  \bibfield  {author} {\bibinfo {author} {\bibfnamefont {E.}~\bibnamefont
  {Barnes}}\ and\ \bibinfo {author} {\bibfnamefont {S.~E.}\ \bibnamefont
  {Economou}},\ }\href {\doibase 10.1103/PhysRevLett.107.047601} {\bibfield
  {journal} {\bibinfo  {journal} {Physical Review Letters}\ }\textbf {\bibinfo
  {volume} {107}},\ \bibinfo {pages} {047601} (\bibinfo {year}
  {2011})}\BibitemShut {NoStop}%
\bibitem [{\citenamefont {Madsen}\ \emph {et~al.}(2011)\citenamefont {Madsen},
  \citenamefont {Ates}, \citenamefont {Lund-Hansen}, \citenamefont {Löffler},
  \citenamefont {Reitzenstein}, \citenamefont {Forchel},\ and\ \citenamefont
  {Lodahl}}]{madsen_observation_2011}%
  \BibitemOpen
  \bibfield  {author} {\bibinfo {author} {\bibfnamefont {K.~H.}\ \bibnamefont
  {Madsen}}, \bibinfo {author} {\bibfnamefont {S.}~\bibnamefont {Ates}},
  \bibinfo {author} {\bibfnamefont {T.}~\bibnamefont {Lund-Hansen}}, \bibinfo
  {author} {\bibfnamefont {A.}~\bibnamefont {Löffler}}, \bibinfo {author}
  {\bibfnamefont {S.}~\bibnamefont {Reitzenstein}}, \bibinfo {author}
  {\bibfnamefont {A.}~\bibnamefont {Forchel}}, \ and\ \bibinfo {author}
  {\bibfnamefont {P.}~\bibnamefont {Lodahl}},\ }\href {\doibase
  10.1103/PhysRevLett.106.233601} {\bibfield  {journal} {\bibinfo  {journal}
  {Physical Review Letters}\ }\textbf {\bibinfo {volume} {106}},\ \bibinfo
  {pages} {233601} (\bibinfo {year} {2011})}\BibitemShut {NoStop}%
\bibitem [{\citenamefont {Urbaszek}\ \emph {et~al.}(2013)\citenamefont
  {Urbaszek}, \citenamefont {Marie}, \citenamefont {Amand}, \citenamefont
  {Krebs}, \citenamefont {Voisin}, \citenamefont {Maletinsky}, \citenamefont
  {Högele},\ and\ \citenamefont {Imamoglu}}]{urbaszek_nuclear_2013}%
  \BibitemOpen
  \bibfield  {author} {\bibinfo {author} {\bibfnamefont {B.}~\bibnamefont
  {Urbaszek}}, \bibinfo {author} {\bibfnamefont {X.}~\bibnamefont {Marie}},
  \bibinfo {author} {\bibfnamefont {T.}~\bibnamefont {Amand}}, \bibinfo
  {author} {\bibfnamefont {O.}~\bibnamefont {Krebs}}, \bibinfo {author}
  {\bibfnamefont {P.}~\bibnamefont {Voisin}}, \bibinfo {author} {\bibfnamefont
  {P.}~\bibnamefont {Maletinsky}}, \bibinfo {author} {\bibfnamefont
  {A.}~\bibnamefont {Högele}}, \ and\ \bibinfo {author} {\bibfnamefont
  {A.}~\bibnamefont {Imamoglu}},\ }\href {\doibase 10.1103/RevModPhys.85.79}
  {\bibfield  {journal} {\bibinfo  {journal} {Reviews of Modern Physics}\
  }\textbf {\bibinfo {volume} {85}},\ \bibinfo {pages} {79} (\bibinfo {year}
  {2013})}\BibitemShut {NoStop}%
\bibitem [{\citenamefont {Economou}\ and\ \citenamefont
  {Barnes}(2014)}]{economou_theory_2014}%
  \BibitemOpen
  \bibfield  {author} {\bibinfo {author} {\bibfnamefont {S.~E.}\ \bibnamefont
  {Economou}}\ and\ \bibinfo {author} {\bibfnamefont {E.}~\bibnamefont
  {Barnes}},\ }\href {\doibase 10.1103/PhysRevB.89.165301} {\bibfield
  {journal} {\bibinfo  {journal} {Physical Review B}\ }\textbf {\bibinfo
  {volume} {89}},\ \bibinfo {pages} {165301} (\bibinfo {year}
  {2014})}\BibitemShut {NoStop}%
\bibitem [{\citenamefont {Munsch}\ \emph {et~al.}(2014)\citenamefont {Munsch},
  \citenamefont {Wuest}, \citenamefont {Kuhlmann}, \citenamefont {Xue},
  \citenamefont {Ludwig}, \citenamefont {Reuter}, \citenamefont {Wieck},
  \citenamefont {Poggio},\ and\ \citenamefont
  {Warburton}}]{munsch_manipulation_2014}%
  \BibitemOpen
  \bibfield  {author} {\bibinfo {author} {\bibfnamefont {M.}~\bibnamefont
  {Munsch}}, \bibinfo {author} {\bibfnamefont {G.}~\bibnamefont {Wuest}},
  \bibinfo {author} {\bibfnamefont {A.~V.}\ \bibnamefont {Kuhlmann}}, \bibinfo
  {author} {\bibfnamefont {F.}~\bibnamefont {Xue}}, \bibinfo {author}
  {\bibfnamefont {A.}~\bibnamefont {Ludwig}}, \bibinfo {author} {\bibfnamefont
  {D.}~\bibnamefont {Reuter}}, \bibinfo {author} {\bibfnamefont {A.~D.}\
  \bibnamefont {Wieck}}, \bibinfo {author} {\bibfnamefont {M.}~\bibnamefont
  {Poggio}}, \ and\ \bibinfo {author} {\bibfnamefont {R.~J.}\ \bibnamefont
  {Warburton}},\ }\href {\doibase 10.1038/nnano.2014.175} {\bibfield  {journal}
  {\bibinfo  {journal} {Nature Nanotechnology}\ }\textbf {\bibinfo {volume}
  {9}},\ \bibinfo {pages} {671} (\bibinfo {year} {2014})}\BibitemShut {NoStop}%
\bibitem [{\citenamefont {Prechtel}\ \emph {et~al.}(2016)\citenamefont
  {Prechtel}, \citenamefont {Kuhlmann}, \citenamefont {Houel}, \citenamefont
  {Ludwig}, \citenamefont {Valentin}, \citenamefont {Wieck},\ and\
  \citenamefont {Warburton}}]{prechtel_decoupling_2016}%
  \BibitemOpen
  \bibfield  {author} {\bibinfo {author} {\bibfnamefont {J.~H.}\ \bibnamefont
  {Prechtel}}, \bibinfo {author} {\bibfnamefont {A.~V.}\ \bibnamefont
  {Kuhlmann}}, \bibinfo {author} {\bibfnamefont {J.}~\bibnamefont {Houel}},
  \bibinfo {author} {\bibfnamefont {A.}~\bibnamefont {Ludwig}}, \bibinfo
  {author} {\bibfnamefont {S.~R.}\ \bibnamefont {Valentin}}, \bibinfo {author}
  {\bibfnamefont {A.~D.}\ \bibnamefont {Wieck}}, \ and\ \bibinfo {author}
  {\bibfnamefont {R.~J.}\ \bibnamefont {Warburton}},\ }\href {\doibase
  10.1038/nmat4704} {\bibfield  {journal} {\bibinfo  {journal} {Nature
  Materials}\ }\textbf {\bibinfo {volume} {15}},\ \bibinfo {pages} {981}
  (\bibinfo {year} {2016})}\BibitemShut {NoStop}%
\bibitem [{\citenamefont {Nutz}\ \emph {et~al.}(2019)\citenamefont {Nutz},
  \citenamefont {Barnes},\ and\ \citenamefont {Economou}}]{nutz_solvable_2019}%
  \BibitemOpen
  \bibfield  {author} {\bibinfo {author} {\bibfnamefont {T.}~\bibnamefont
  {Nutz}}, \bibinfo {author} {\bibfnamefont {E.}~\bibnamefont {Barnes}}, \ and\
  \bibinfo {author} {\bibfnamefont {S.~E.}\ \bibnamefont {Economou}},\ }\href
  {\doibase 10.1103/PhysRevB.99.035439} {\bibfield  {journal} {\bibinfo
  {journal} {Physical Review B}\ }\textbf {\bibinfo {volume} {99}},\ \bibinfo
  {pages} {035439} (\bibinfo {year} {2019})}\BibitemShut {NoStop}%
\bibitem [{\citenamefont {Androvitsaneas}\ \emph {et~al.}(2016)\citenamefont
  {Androvitsaneas}, \citenamefont {Young}, \citenamefont {Schneider},
  \citenamefont {Maier}, \citenamefont {Kamp}, \citenamefont {Hoefling},
  \citenamefont {Knauer}, \citenamefont {Harbord}, \citenamefont {Hu},
  \citenamefont {Rarity},\ and\ \citenamefont
  {Oulton}}]{androvitsaneas_charged_2016}%
  \BibitemOpen
  \bibfield  {author} {\bibinfo {author} {\bibfnamefont {P.}~\bibnamefont
  {Androvitsaneas}}, \bibinfo {author} {\bibfnamefont {A.~B.}\ \bibnamefont
  {Young}}, \bibinfo {author} {\bibfnamefont {C.}~\bibnamefont {Schneider}},
  \bibinfo {author} {\bibfnamefont {S.}~\bibnamefont {Maier}}, \bibinfo
  {author} {\bibfnamefont {M.}~\bibnamefont {Kamp}}, \bibinfo {author}
  {\bibfnamefont {S.}~\bibnamefont {Hoefling}}, \bibinfo {author}
  {\bibfnamefont {S.}~\bibnamefont {Knauer}}, \bibinfo {author} {\bibfnamefont
  {E.}~\bibnamefont {Harbord}}, \bibinfo {author} {\bibfnamefont {C.~Y.}\
  \bibnamefont {Hu}}, \bibinfo {author} {\bibfnamefont {J.~G.}\ \bibnamefont
  {Rarity}}, \ and\ \bibinfo {author} {\bibfnamefont {R.}~\bibnamefont
  {Oulton}},\ }\href {\doibase 10.1103/PhysRevB.93.241409} {\bibfield
  {journal} {\bibinfo  {journal} {Physical Review B}\ }\textbf {\bibinfo
  {volume} {93}},\ \bibinfo {pages} {241409} (\bibinfo {year}
  {2016})}\BibitemShut {NoStop}%
\bibitem [{\citenamefont {Auffèves-Garnier}\ \emph {et~al.}(2007)\citenamefont
  {Auffèves-Garnier}, \citenamefont {Simon}, \citenamefont {Gérard},\ and\
  \citenamefont {Poizat}}]{auffeves-garnier_giant_2007}%
  \BibitemOpen
  \bibfield  {author} {\bibinfo {author} {\bibfnamefont {A.}~\bibnamefont
  {Auffèves-Garnier}}, \bibinfo {author} {\bibfnamefont {C.}~\bibnamefont
  {Simon}}, \bibinfo {author} {\bibfnamefont {J.-M.}\ \bibnamefont {Gérard}},
  \ and\ \bibinfo {author} {\bibfnamefont {J.-P.}\ \bibnamefont {Poizat}},\
  }\href {\doibase 10.1103/PhysRevA.75.053823} {\bibfield  {journal} {\bibinfo
  {journal} {Physical Review A}\ }\textbf {\bibinfo {volume} {75}},\ \bibinfo
  {pages} {053823} (\bibinfo {year} {2007})}\BibitemShut {NoStop}%
\bibitem [{\citenamefont {Abragam}(1961)}]{abragam_principles_1961}%
  \BibitemOpen
  \bibfield  {author} {\bibinfo {author} {\bibfnamefont {A.}~\bibnamefont
  {Abragam}},\ }\href@noop {} {\emph {\bibinfo {title} {The {Principles} of
  {Nuclear} {Magnetism}}}}\ (\bibinfo  {publisher} {Clarendon Press},\ \bibinfo
  {address} {Oxford},\ \bibinfo {year} {1961})\BibitemShut {NoStop}%
\bibitem [{\citenamefont {Androvitsaneas}\ \emph {et~al.}(2019)\citenamefont
  {Androvitsaneas}, \citenamefont {Young}, \citenamefont {Lennon},
  \citenamefont {Schneider}, \citenamefont {Maier}, \citenamefont {Hinchliff},
  \citenamefont {Atkinson}, \citenamefont {Harbord}, \citenamefont {Kamp},
  \citenamefont {Hoefling}, \citenamefont {Rarity},\ and\ \citenamefont
  {Oulton}}]{androvitsaneas_efficient_2019}%
  \BibitemOpen
  \bibfield  {author} {\bibinfo {author} {\bibfnamefont {P.}~\bibnamefont
  {Androvitsaneas}}, \bibinfo {author} {\bibfnamefont {A.}~\bibnamefont
  {Young}}, \bibinfo {author} {\bibfnamefont {J.}~\bibnamefont {Lennon}},
  \bibinfo {author} {\bibfnamefont {C.}~\bibnamefont {Schneider}}, \bibinfo
  {author} {\bibfnamefont {S.}~\bibnamefont {Maier}}, \bibinfo {author}
  {\bibfnamefont {J.}~\bibnamefont {Hinchliff}}, \bibinfo {author}
  {\bibfnamefont {G.}~\bibnamefont {Atkinson}}, \bibinfo {author}
  {\bibfnamefont {E.}~\bibnamefont {Harbord}}, \bibinfo {author} {\bibfnamefont
  {M.}~\bibnamefont {Kamp}}, \bibinfo {author} {\bibfnamefont {S.}~\bibnamefont
  {Hoefling}}, \bibinfo {author} {\bibfnamefont {J.~G.}\ \bibnamefont
  {Rarity}}, \ and\ \bibinfo {author} {\bibfnamefont {R.}~\bibnamefont
  {Oulton}},\ }\href {\doibase 10.1021/acsphotonics.8b01380} {\bibfield
  {journal} {\bibinfo  {journal} {ACS Photonics}\ } (\bibinfo {year} {2019}),\
  10.1021/acsphotonics.8b01380}\BibitemShut {NoStop}%
\bibitem [{\citenamefont {Misra}\ and\ \citenamefont
  {Sudarshan}(1977)}]{misra_zenos_1977}%
  \BibitemOpen
  \bibfield  {author} {\bibinfo {author} {\bibfnamefont {B.}~\bibnamefont
  {Misra}}\ and\ \bibinfo {author} {\bibfnamefont {E.~C.~G.}\ \bibnamefont
  {Sudarshan}},\ }\href {\doibase 10.1063/1.523304} {\bibfield  {journal}
  {\bibinfo  {journal} {Journal of Mathematical Physics}\ }\textbf {\bibinfo
  {volume} {18}},\ \bibinfo {pages} {756} (\bibinfo {year} {1977})}\BibitemShut
  {NoStop}%
\bibitem [{\citenamefont {Itano}\ \emph {et~al.}(1990)\citenamefont {Itano},
  \citenamefont {Heinzen}, \citenamefont {Bollinger},\ and\ \citenamefont
  {Wineland}}]{itano_quantum_1990}%
  \BibitemOpen
  \bibfield  {author} {\bibinfo {author} {\bibfnamefont {W.~M.}\ \bibnamefont
  {Itano}}, \bibinfo {author} {\bibfnamefont {D.~J.}\ \bibnamefont {Heinzen}},
  \bibinfo {author} {\bibfnamefont {J.~J.}\ \bibnamefont {Bollinger}}, \ and\
  \bibinfo {author} {\bibfnamefont {D.~J.}\ \bibnamefont {Wineland}},\ }\href
  {\doibase 10.1103/PhysRevA.41.2295} {\bibfield  {journal} {\bibinfo
  {journal} {Physical Review A}\ }\textbf {\bibinfo {volume} {41}},\ \bibinfo
  {pages} {2295} (\bibinfo {year} {1990})}\BibitemShut {NoStop}%
\bibitem [{\citenamefont {Block}\ and\ \citenamefont
  {Berman}(1991)}]{block_quantum_1991}%
  \BibitemOpen
  \bibfield  {author} {\bibinfo {author} {\bibfnamefont {E.}~\bibnamefont
  {Block}}\ and\ \bibinfo {author} {\bibfnamefont {P.~R.}\ \bibnamefont
  {Berman}},\ }\href {\doibase 10.1103/PhysRevA.44.1466} {\bibfield  {journal}
  {\bibinfo  {journal} {Physical Review A}\ }\textbf {\bibinfo {volume} {44}},\
  \bibinfo {pages} {1466} (\bibinfo {year} {1991})}\BibitemShut {NoStop}%
\bibitem [{\citenamefont {Facchi}\ \emph {et~al.}(2000)\citenamefont {Facchi},
  \citenamefont {Gorini}, \citenamefont {Marmo}, \citenamefont {Pascazio},\
  and\ \citenamefont {Sudarshan}}]{facchi_quantum_2000}%
  \BibitemOpen
  \bibfield  {author} {\bibinfo {author} {\bibfnamefont {P.}~\bibnamefont
  {Facchi}}, \bibinfo {author} {\bibfnamefont {V.}~\bibnamefont {Gorini}},
  \bibinfo {author} {\bibfnamefont {G.}~\bibnamefont {Marmo}}, \bibinfo
  {author} {\bibfnamefont {S.}~\bibnamefont {Pascazio}}, \ and\ \bibinfo
  {author} {\bibfnamefont {E.}~\bibnamefont {Sudarshan}},\ }\href {\doibase
  10.1016/S0375-9601(00)00566-1} {\bibfield  {journal} {\bibinfo  {journal}
  {Physics Letters A}\ }\textbf {\bibinfo {volume} {275}},\ \bibinfo {pages}
  {12} (\bibinfo {year} {2000})}\BibitemShut {NoStop}%
\bibitem [{\citenamefont {Pascazio}(2014)}]{pascazio_all_2014}%
  \BibitemOpen
  \bibfield  {author} {\bibinfo {author} {\bibfnamefont {S.}~\bibnamefont
  {Pascazio}},\ }\href
  {http://www.worldscientific.com/doi/pdf/10.1142/S1230161214400071} {\bibfield
   {journal} {\bibinfo  {journal} {Open Systems \& Information Dynamics}\
  }\textbf {\bibinfo {volume} {21}},\ \bibinfo {pages} {1440007} (\bibinfo
  {year} {2014})}\BibitemShut {NoStop}%
\bibitem [{\citenamefont {Zhang}\ and\ \citenamefont
  {Fan}(2015)}]{zhang_zeno_2015}%
  \BibitemOpen
  \bibfield  {author} {\bibinfo {author} {\bibfnamefont {Y.-R.}\ \bibnamefont
  {Zhang}}\ and\ \bibinfo {author} {\bibfnamefont {H.}~\bibnamefont {Fan}},\
  }\href {\doibase 10.1038/srep11509} {\bibfield  {journal} {\bibinfo
  {journal} {Scientific Reports}\ }\textbf {\bibinfo {volume} {5}},\ \bibinfo
  {pages} {11509} (\bibinfo {year} {2015})}\BibitemShut {NoStop}%
\bibitem [{\citenamefont {Christensen}\ \emph {et~al.}(2018)\citenamefont
  {Christensen}, \citenamefont {Iles-Smith}, \citenamefont {Petersen},
  \citenamefont {M{\o}rk},\ and\ \citenamefont
  {McCutcheon}}]{christensen_driving-induced_2018}%
  \BibitemOpen
  \bibfield  {author} {\bibinfo {author} {\bibfnamefont {C.~N.}\ \bibnamefont
  {Christensen}}, \bibinfo {author} {\bibfnamefont {J.}~\bibnamefont
  {Iles-Smith}}, \bibinfo {author} {\bibfnamefont {T.~S.}\ \bibnamefont
  {Petersen}}, \bibinfo {author} {\bibfnamefont {J.}~\bibnamefont {M{\o}rk}}, \
  and\ \bibinfo {author} {\bibfnamefont {D.~P.~S.}\ \bibnamefont
  {McCutcheon}},\ }\href {\doibase 10.1103/PhysRevA.97.063807} {\bibfield
  {journal} {\bibinfo  {journal} {Physical Review A}\ }\textbf {\bibinfo
  {volume} {97}},\ \bibinfo {pages} {063807} (\bibinfo {year}
  {2018})}\BibitemShut {NoStop}%
\bibitem [{\citenamefont {Gardiner}\ and\ \citenamefont
  {Collett}(1985)}]{gardiner_input_1985}%
  \BibitemOpen
  \bibfield  {author} {\bibinfo {author} {\bibfnamefont {C.~W.}\ \bibnamefont
  {Gardiner}}\ and\ \bibinfo {author} {\bibfnamefont {M.~J.}\ \bibnamefont
  {Collett}},\ }\href {\doibase 10.1103/PhysRevA.31.3761} {\bibfield  {journal}
  {\bibinfo  {journal} {Physical Review A}\ }\textbf {\bibinfo {volume} {31}},\
  \bibinfo {pages} {3761} (\bibinfo {year} {1985})}\BibitemShut {NoStop}%
\bibitem [{\citenamefont {Walls}\ and\ \citenamefont
  {Milburn}(2008)}]{walls_quantum_2008}%
  \BibitemOpen
  \bibfield  {author} {\bibinfo {author} {\bibfnamefont {D.~F.}\ \bibnamefont
  {Walls}}\ and\ \bibinfo {author} {\bibfnamefont {G.~J.}\ \bibnamefont
  {Milburn}},\ }\href {//www.springer.com/gb/book/9783540285731} {\emph
  {\bibinfo {title} {Quantum {Optics}}}},\ \bibinfo {edition} {2nd}\ ed.\
  (\bibinfo  {publisher} {Springer-Verlag},\ \bibinfo {address} {Berlin
  Heidelberg},\ \bibinfo {year} {2008})\BibitemShut {NoStop}%
\bibitem [{\citenamefont {Klauser}\ \emph {et~al.}(2008)\citenamefont
  {Klauser}, \citenamefont {Coish},\ and\ \citenamefont
  {Loss}}]{klauser_nuclear_2008}%
  \BibitemOpen
  \bibfield  {author} {\bibinfo {author} {\bibfnamefont {D.}~\bibnamefont
  {Klauser}}, \bibinfo {author} {\bibfnamefont {W.~A.}\ \bibnamefont {Coish}},
  \ and\ \bibinfo {author} {\bibfnamefont {D.}~\bibnamefont {Loss}},\ }\href
  {http://journals.aps.org/prb/abstract/10.1103/PhysRevB.78.205301} {\bibfield
  {journal} {\bibinfo  {journal} {Physical Review B}\ }\textbf {\bibinfo
  {volume} {78}},\ \bibinfo {pages} {205301} (\bibinfo {year}
  {2008})}\BibitemShut {NoStop}%
\bibitem [{\citenamefont {Plenio}\ and\ \citenamefont
  {Knight}(1998)}]{plenio_quantum-jump_1998}%
  \BibitemOpen
  \bibfield  {author} {\bibinfo {author} {\bibfnamefont {M.~B.}\ \bibnamefont
  {Plenio}}\ and\ \bibinfo {author} {\bibfnamefont {P.~L.}\ \bibnamefont
  {Knight}},\ }\href {\doibase 10.1103/RevModPhys.70.101} {\bibfield  {journal}
  {\bibinfo  {journal} {Reviews of Modern Physics}\ }\textbf {\bibinfo {volume}
  {70}},\ \bibinfo {pages} {101} (\bibinfo {year} {1998})}\BibitemShut
  {NoStop}%
\bibitem [{\citenamefont {Carmichael}\ \emph {et~al.}(1989)\citenamefont
  {Carmichael}, \citenamefont {Singh}, \citenamefont {Vyas},\ and\
  \citenamefont {Rice}}]{carmichael_photoelectron_1989}%
  \BibitemOpen
  \bibfield  {author} {\bibinfo {author} {\bibfnamefont {H.~J.}\ \bibnamefont
  {Carmichael}}, \bibinfo {author} {\bibfnamefont {S.}~\bibnamefont {Singh}},
  \bibinfo {author} {\bibfnamefont {R.}~\bibnamefont {Vyas}}, \ and\ \bibinfo
  {author} {\bibfnamefont {P.~R.}\ \bibnamefont {Rice}},\ }\href {\doibase
  10.1103/PhysRevA.39.1200} {\bibfield  {journal} {\bibinfo  {journal}
  {Physical Review A}\ }\textbf {\bibinfo {volume} {39}},\ \bibinfo {pages}
  {1200} (\bibinfo {year} {1989})}\BibitemShut {NoStop}%
\bibitem [{\citenamefont {Nazir}\ and\ \citenamefont
  {McCutcheon}(2016)}]{nazir_modelling_2016}%
  \BibitemOpen
  \bibfield  {author} {\bibinfo {author} {\bibfnamefont {A.}~\bibnamefont
  {Nazir}}\ and\ \bibinfo {author} {\bibfnamefont {D.~P.~S.}\ \bibnamefont
  {McCutcheon}},\ }\href {\doibase 10.1088/0953-8984/28/10/103002} {\bibfield
  {journal} {\bibinfo  {journal} {Journal of Physics: Condensed Matter}\
  }\textbf {\bibinfo {volume} {28}},\ \bibinfo {pages} {103002} (\bibinfo
  {year} {2016})}\BibitemShut {NoStop}%
\bibitem [{\citenamefont {Cywiński}\ \emph {et~al.}(2009)\citenamefont
  {Cywiński}, \citenamefont {Witzel},\ and\ \citenamefont
  {Sarma}}]{cywinski_pure_2009}%
  \BibitemOpen
  \bibfield  {author} {\bibinfo {author} {\bibfnamefont {L.}~\bibnamefont
  {Cywiński}}, \bibinfo {author} {\bibfnamefont {W.~M.}\ \bibnamefont
  {Witzel}}, \ and\ \bibinfo {author} {\bibfnamefont {S.~D.}\ \bibnamefont
  {Sarma}},\ }\href
  {http://journals.aps.org/prb/abstract/10.1103/PhysRevB.79.245314} {\bibfield
  {journal} {\bibinfo  {journal} {Physical Review B}\ }\textbf {\bibinfo
  {volume} {79}},\ \bibinfo {pages} {245314} (\bibinfo {year}
  {2009})}\BibitemShut {NoStop}%
\bibitem [{\citenamefont {Latta}\ \emph {et~al.}(2009)\citenamefont {Latta},
  \citenamefont {H{\"o}gele}, \citenamefont {Zhao}, \citenamefont {Vamivakas},
  \citenamefont {Maletinsky}, \citenamefont {Kroner}, \citenamefont {Dreiser},
  \citenamefont {Carusotto}, \citenamefont {Badolato}, \citenamefont {Schuh},
  \citenamefont {Wegscheider}, \citenamefont {Atature},\ and\ \citenamefont
  {Imamoglu}}]{latta_confluence_2009}%
  \BibitemOpen
  \bibfield  {author} {\bibinfo {author} {\bibfnamefont {C.}~\bibnamefont
  {Latta}}, \bibinfo {author} {\bibfnamefont {A.}~\bibnamefont {H{\"o}gele}},
  \bibinfo {author} {\bibfnamefont {Y.}~\bibnamefont {Zhao}}, \bibinfo {author}
  {\bibfnamefont {A.~N.}\ \bibnamefont {Vamivakas}}, \bibinfo {author}
  {\bibfnamefont {P.}~\bibnamefont {Maletinsky}}, \bibinfo {author}
  {\bibfnamefont {M.}~\bibnamefont {Kroner}}, \bibinfo {author} {\bibfnamefont
  {J.}~\bibnamefont {Dreiser}}, \bibinfo {author} {\bibfnamefont
  {I.}~\bibnamefont {Carusotto}}, \bibinfo {author} {\bibfnamefont
  {A.}~\bibnamefont {Badolato}}, \bibinfo {author} {\bibfnamefont
  {D.}~\bibnamefont {Schuh}}, \bibinfo {author} {\bibfnamefont
  {W.}~\bibnamefont {Wegscheider}}, \bibinfo {author} {\bibfnamefont
  {M.}~\bibnamefont {Atature}}, \ and\ \bibinfo {author} {\bibfnamefont
  {A.}~\bibnamefont {Imamoglu}},\ }\href {\doibase 10.1038/nphys1363}
  {\bibfield  {journal} {\bibinfo  {journal} {Nature Physics}\ }\textbf
  {\bibinfo {volume} {5}},\ \bibinfo {pages} {758} (\bibinfo {year}
  {2009})}\BibitemShut {NoStop}%
\bibitem [{\citenamefont {Dreiser}\ \emph {et~al.}(2008)\citenamefont
  {Dreiser}, \citenamefont {Atat{\"u}re}, \citenamefont {Galland},
  \citenamefont {M{\"u}ller}, \citenamefont {Badolato},\ and\ \citenamefont
  {Imamoglu}}]{dreiser_optical_2008}%
  \BibitemOpen
  \bibfield  {author} {\bibinfo {author} {\bibfnamefont {J.}~\bibnamefont
  {Dreiser}}, \bibinfo {author} {\bibfnamefont {M.}~\bibnamefont
  {Atat{\"u}re}}, \bibinfo {author} {\bibfnamefont {C.}~\bibnamefont
  {Galland}}, \bibinfo {author} {\bibfnamefont {T.}~\bibnamefont {M{\"u}ller}},
  \bibinfo {author} {\bibfnamefont {A.}~\bibnamefont {Badolato}}, \ and\
  \bibinfo {author} {\bibfnamefont {A.}~\bibnamefont {Imamoglu}},\ }\href
  {\doibase 10.1103/PhysRevB.77.075317} {\bibfield  {journal} {\bibinfo
  {journal} {Physical Review B}\ }\textbf {\bibinfo {volume} {77}},\ \bibinfo
  {pages} {075317} (\bibinfo {year} {2008})}\BibitemShut {NoStop}%
\end{thebibliography}%
\end{document}